\documentclass[namedreferences]{solarphysics}
\usepackage[optionalrh,solaromanenum]{spr-sola-addons} 
\usepackage{epsfig}                     
\usepackage{graphicx}                    
\usepackage{color}                       
\usepackage{hyperref}
\usepackage{natbib}
\usepackage{solaheader}

\usepackage{times}
\usepackage{txfonts}
\usepackage{multirow}
\usepackage{rotating}

\begin{document}

\begin{article}

\begin{opening}

\title{A Helioseismic Perspective on the Depth of the Minimum Between Solar Cycles 23 and 24}

\author{A.-M.~\surname{Broomhall}$^{1,2}$}

%
  \institute{$^1$Institute of Advanced Studies, University of Warwick, Coventry, CV4 7HS, UK\\
                    email: \url{a-m.broomhall@warwick.ac.uk}\\
                $^2$Centre for Fusion, Space, and Astrophysics, Department of Physics, University of Warwick, Coventry CV4 7AL, UK}

%
\runningauthor{A.-M. Broomhall}
\runningtitle{Helioseismic Depth of Cycle Minimum}

\begin{abstract}
The solar-activity-cycle minimum observed between Cycles 23 and 24 is generally regarded as being unusually deep and long. That minimum is being followed by one of the smallest amplitude cycles in recent history. We perform an in-depth analysis of this minimum with helioseismology. We use Global Oscillation Network Group (GONG) data to demonstrate that the frequencies of helioseismic oscillations are a sensitive probe of the Sun's magnetic field: The frequencies of the helioseismic oscillations were found to be systematically lower in the minimum following Cycle 23 than in the minimum preceding it. This difference is statistically significant and may indicate that the Sun's global magnetic field was weaker in the minimum following Cycle 23. The size of the shift in oscillation frequencies between the two minima is dependent on the frequency of the oscillation and takes the same functional form as the frequency dependence observed when the frequencies at cycle maximum are compared with the cycle-minimum frequencies. This implies that the same near-surface magnetic perturbation is responsible. Finally, we determine that the difference in the mean magnetic field between the minimum preceding Cycle 23 and that following it is approximately 1G. 

\end{abstract}

%
\keywords{Helioseismology, Observations; Oscillations, Solar; Solar Cycle, Observations}

\end{opening}

\section{Introduction}\label{section[introduction]}
The Sun's magnetic-activity cycle varies primarily on a time scale of 11years. While much attention is paid to the dynamic features primarily observed close to activity maxima, activity minima can be interesting in their own right. The activity minimum between Cycles 23 and 24 is regarded as being unusually long and deep. For example, 2008 had more spotless days than any year since 1913 \citep{2010ASPC..428....3S}, the lowest ever yearly averaged value of the 10.7\,cm flux was recorded \citep[since systematic measurements began in 1947;][]{2013JPhCS.440a2001B}, and the solar-wind pressure dropped to a 50-year low \citep[\textit{e.g.}][]{2008GeoRL..3518103M}.  Many measures of solar activity are limited by the fact that they are predominantly sensitive to strong active regions and so are positive definite, which sets a hard lower limit on the activity measures: the values of sunspot area and sunspot number cannot have negative values, which restricts what sunspot proxies can reveal about the absolute depths of the minima. 

Minima can be informative for models of solar dynamos. \citet{2010GeoRL..37.6104D} demonstrated that the length of a cycle minimum, as seen in sunspot data, is anticorrelated with the depth of the minimum and speculate that the extended period in which oppositely directed toroidal bands of magnetic field either side of the equator lie in close proximity to each other leads to flux annihilation in the solar interior and, therefore, fewer surface manifestations of magnetic activity. Furthermore, \citet{2010GeoRL..37.6104D} state that such cancellation is common to all dynamo models that include shearing as a mechanism for generating spot producing toroidal fields \citep[\textit{e.g.}][]{1991ApJ...375..761W, 1999ApJ...518..508D}. The importance of shearing in the context of stellar dynamos has recently been brought into question by the results of \citet{2016Natur.535..526W}, who found that low-mass fully convective stars obey the same rotation--activity relation as those stars that switch from radiation to convection in their interiors. 

Despite the link between the length and depth of a minimum, \citet{2010GeoRL..37.6104D} find that the length of a cycle minimum cannot be used to predict the amplitude of the next cycle.  Polar fields during cycle minima, on the other hand, are generally regarded as a good indicator of the strength of the upcoming activity Cycle \citep[e.g.][]{2009ApJ...694L..11W}. Polar fields were observed to be weaker during the minimum following Cycle 23 than the minimum preceding it \citep[\textit{e.g.}][]{2010LRSP....7....1H, 2011SoPh..274..195D, 2012ApJ...750L..42G}, while Cycle 24 is far weaker than Cycle 23, thus supporting the predictive nature of the polar fields. 

Helioseismology uses the Sun's natural resonant oscillations to probe the solar interior and the dominant helioseismic oscillations are acoustic \textit{p}-modes. The observed change in frequency as a function of time of these \textit{p}-modes [$\delta\nu$] can be used as a proxy of the near-surface internal magnetic field \citep[see][for a recent review]{2014SSRv..186..191B}.  The helioseismic frequency shift is sensitive to both the strong and the weak components of the magnetic field \citep[\textit{e.g.}][]{2007ApJ...659.1749C, 2009ApJ...695.1567J, 2015SoPh..290.3095B} and as such should provide a good measure of the depth of the last solar minimum. Helioseismology has already been used to demonstrate that the last minimum differed from the one preceding it: \citet{2010ApJ...711L..84T} and \citet{2011ApJ...739....6J}, for example, show that the minimum following Cycle 23 was deeper than the minimum preceding it. \citet{2010ApJ...720..494A} demonstrated that the high-latitude meridional flow was slower in the most recent minimum, while \citet{2009ApJ...701L..87H} determined that the equatorward branch of the torsional oscillation was noticeably slower to move toward the Equator during the minimum following Cycle 23 than the minimum preceding it. \citet{2011JPhCS.271a2074H} demonstrated that the high-latitude poleward-propagating band of faster-than-average rotation, which was clearly visible in the minimum preceding Cycle 23, was not detected in the most recent minimum, leading to speculations that Cycle 25 may be weak or non-existent. While \citet{2013ApJ...767L..20H} showed the high-latitude flow eventually became visible, this flow was significantly weaker than that observed in the previous cycle. They speculate that this may be in response to the above-mentioned weaker polar fields.

Helioseismology has provided useful insights into the long, deep minimum that preceded Cycle 24 specifically. For example, a number of studies have demonstrated that the \textit{p}-mode minimum was significantly deeper than the minimum observed in surface-activity proxies \citep[\textit{e.g.}][]{2009ApJ...700L.162B, 2010ApJ...711L..84T, 2011ApJ...739....6J, 2015MNRAS.454.4120H}. This could indicate a change in relationship between the activity proxies and the helioseismic frequencies: Both \citet{2010ApJ...711L..84T} and \citet{2013SoPh..282....1T} found the frequency shifts and activity proxies were anti-correlated in the minimum following Cycle 23 using intermediate and high-degree modes respectively. \citet{2011ApJ...739....6J} further demonstrated that a double-dip minimum was observed in the \textit{p}-mode frequencies and that the timing of the minimum depended on mode degree, while \citet{2016ApJ...828...41S} established that the timing of the minimum depends on the latitudinal sensitivity of the modes under consideration and, to a lesser extent, mode frequency. There is also evidence for hemispheric asymmetry in helioseismic-activity proxies: \citet{2015ApJ...812...20T} showed that the minimum phase lasted longer in the northern hemisphere than the southern and \citet{2016ApJ...828...41S} determined that there is a hemispheric dependence in the timing of the minimum, with the southern hemisphere delayed compared to the North by approximately one year. 

Here, we move away from studying the timing and temporal structure of the minima and instead perform a direct comparison of the \textit{p}-mode frequencies observed in the two minima that encompass Cycle 23. Not only does this allow us to determine directly which minima was deeper (including the statistical significance of the difference) but it will also allow a determination of the frequency dependence of the \textit{p}-mode frequency shifts, which is known to relate to the depth dependence of the perturbation \citep[\textit{e.g.}][]{1990LNP...367..283G, 1990Natur.345..779L, 1991ApJ...370..752G, 2005ApJ...625..548D}. We begin, in Section \ref{section[data]}, by describing the data used in this analysis. Section \ref{section[proxies]} contains a discussion of atmospheric measures of solar activity as seen in the last two solar minima. The main results of the helioseismic analyses are given in Section \ref{section[results]}. In Section \ref{section[discussion]} we discuss the wider implications of these results.

\section{Data}\label{section[data]}

The Global Oscillation Network Group (GONG) makes resolved Doppler-velocity observations of the Sun. The GONG standard pipeline determines mode frequencies for oscillations with harmonic degrees in the range $0\le\ell\le150$. However, here we focus on those modes in the range $11\le\ell\le150$ as the frequencies of the lower-$\ell$ modes are more uncertain. Frequencies were determined using the standard GONG 108-day time series, with a start-time separation of 36\,days and the frequencies can be obtained from \href{http://gong2.nso.edu/archive/patch.pl?menutype=s}{\textsf{gong2.nso.edu}}.  Here we used the \textsf{mrv1y} files, which give the Clebsch--Gordan rotation coefficients for each combination of harmonic degree [$\ell$] and radial order [$n$] for which there are sufficient data. The power spectra are fitted with orthogonal polynomials as defined by \citet{1991ApJ...369..557R}: \begin{equation}\label{eq[CG]}
\nu_{n,\ell,m}=\nu_{n,\ell}+\sum_i c_{i,n,\ell}\gamma_{i,\ell}(m),
\end{equation}
where $\gamma_{i,\ell}(m)$ are the orthogonal polynomials for a given value of $\ell$ as defined by \citet{1991ApJ...369..557R}, and $c_{i,n,\ell}$ are the Clebsch--Gordan coefficients. We considered only the values of $\nu_{n,\ell}$ given in the tables, and in the remainder of the article we refer to these as the $m$-averaged frequencies. Here we considered oscillations with the largest amplitudes, which corresponds to oscillations with $m$-averaged frequencies in the range $1700\le\nu_{n,\ell}\le4000\,\rm\mu Hz$.

\begin{figure*}
  \centering
  \includegraphics[clip, width=0.7\textwidth]{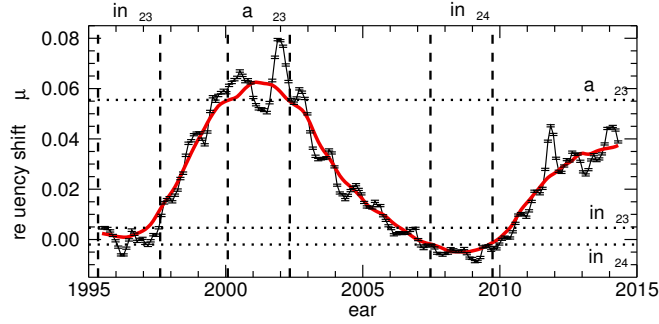}\\
  \caption{Mean frequency shift as a function of time, seen in \textit{p}-modes with $11\le\ell\le150$ and $1700\le\nu_{n,\ell}\le4000\,\rm\mu Hz$ (black data with uncertainties). Also plotted are the smoothed frequency shifts, where smoothing was performed with a 25-point boxcar, \textit{i.e.} smoothing was performed over approximately 2.5\,years (red solid line with no measurement uncertainties). Note that edge effects of the smoothing are visible at the start and ends of the smoothed data. The vertical-dashed lines indicate the dates defined as $\rm Min_{23}$, $\rm Max_{23}$, and $\rm Min_{24}$ respectively, while the horizontal-dotted lines show the frequency shifts used to define these periods (see Section \ref{section[data]}). }
  \label{figure[freq_shift_vs_time]}
\end{figure*}

The helioseismic data were used to define the respective minima and maxima. Figure \ref{figure[freq_shift_vs_time]} shows the weighted-average frequency shifts as a function of time. The frequency shifts were defined in comparison to the minimum preceding Cycle 23 and the weights were taken to be the formal uncertainties in the mode frequencies. We note that it appears even in Figure \ref{figure[freq_shift_vs_time]} that the minimum following Cycle 23 was deeper than the minimum preceding Cycle 23, in agreement with \citet{2010ApJ...711L..84T, 2011ApJ...739....6J}. 

Figure \ref{figure[freq_shift_vs_time]} shows that while the frequency shifts vary primarily on an 11-year time\-scale, they also show evidence for shorter-timescale variations. These shorter timescales are often referred to as the quasi-biennial oscillation \citep[QBO; \textit{e.g.}][]{2009ApJ...700L.162B, 2012MNRAS.420.1405B, 2012A&A...539A.135S, 2013ApJ...765..100S}. Furthermore, \citet{2011ApJ...739....6J} demonstrated that, for certain modes, the minimum following Cycle 23, which is the primary focus of this study, actually consisted of a double minimum, while for other modes only a single minimum was observed. The distinction was determined by the latitudinal sensitivity of the oscillations, with those sensitive to higher latitudes showing a double minimum, while those only sensitive to mid- to low-latitudes displaying a single minimum. \citet{2011ApJ...739....6J} also demonstrate the appearance of a double-minimum depends on the lower turning point of the oscillations, with \textit{p}-modes that travel to the radiative interior and core showing a double-dip minimum, while those trapped in the convection zone only show a single minimum. A double minimum is clearly visible in the frequency shifts plotted in Figure \ref{figure[freq_shift_vs_time]}, and we note that the minimum preceding Cycle 23 also shows a double-dip structure. In order to define the minimum epoch we therefore smoothed the frequency shifts, where smoothing was performed with a 25-point boxcar, \textit{i.e.} over approximately 2.5\,years.

The minimum following Cycle 23 was taken to be between  21 June 2007 and 25 November 2009 inclusive. For the remainder of the article we refer to this minimum as $\rm Min_{24}$. The definition of the minimum was based upon the smoothed average frequency shift as a function of time plotted in Figure \ref{figure[freq_shift_vs_time]} and corresponds to the epoch where the smoothed frequency shift was below $-0.002\,\rm\mu Hz$. We note that it has been shown that the timing of the minimum depends on the properties of the modes under consideration \citep{2011ApJ...739....6J, 2015ApJ...812...20T, 2016ApJ...828...41S}. However, this epoch covers the minimum as identified by these and other authors \citep[\textit{e.g.}][]{2010ApJ...711L..84T, 2013SoPh..282....1T}, with the exception of those modes only sensitive to latitudes below $15^\circ$ \citep{2016ApJ...828...41S}. This epoch contains 21 GONG data sets. 

We then defined the minimum preceding Cycle 23 using the same frequency shifts as a function of time. Despite the fact that the minimum following Cycle 23 was generally accepted to be longer than the minimum preceding Cycle 23, for consistency, we ensured that the minimum preceding Cycle 23 also contained 21 GONG data sets. Since the smoothed frequency shifts suffer edge effects this close to the start of the data set, we used the raw frequency shifts to define this minimum: The minimum preceding Cycle 23 corresponds to the epoch when the unsmoothed frequency shifts were continuously below $0.00465\,\rm\mu Hz$. The minimum preceding Cycle 23 was, therefore, taken to be between 7 May 1995 and 11 August 1997 inclusive. For the remainder of the article we refer to this minimum as $\rm Min_{23}$. As with $\rm Min_{24}$ this encompasses other definitions of the minimum, with the exception of the low-latitude modes examined by \citet{2016ApJ...828...41S}. Only modes present in all of the GONG frequency sets in both $\rm Min_{23}$ and $\rm Min_{24}$ were used in the analysis, and 1551 modes were considered in total. 

We also compare the minimum preceding Cycle 23 with the maximum of Cycle 23. Again an epoch of the same length was considered, containing 21 GONG data sets and covering the period between 29 January 2000 and 5  May 2002. For the remainder of the article we refer to this epoch as $\rm Max_{23}$. As with $\rm Min_{24}$, this was defined using the average smoothed frequency shifts as a function of time and corresponds to the times when the smoothed frequency shifts were greater than $0.0555\,\rm\mu Hz$.  Again we only consider modes observed in all time series during both $\rm Min_{23}$ and $\rm Max_{23}$ and so for this analysis 1541 modes were included.

For comparison purposes we also examine a number of alternative activity proxies. We consider NOAA 
International Sunspot Number (\href{http://www.ngdc.noaa.gov/stp/space-weather/solar-data/solar-
indices/sunspot-numbers/international/}{\textsf{www.ngdc.noaa.gov}}: SSN: see discussion by \opencite{2014SSRv..186...35C}), sunspot area (\href{http://solarscience.msfc.nasa.gov/greenwch.shtml}{\textsf{solarscience.msfc.nasa.gov}}: SSA: \opencite{2010LRSP....7....1H}), Mount Wilson Sunspot Index (\href{http://obs.astro.ucla.edu/150_data.HTML}{\textsf{obs.astro.ucla.edu}}: MWSI), Magnetic Plage Strength Index (\href{http://obs.astro.ucla.edu/150_data.html}{\textsf{obs.astro.ucla.edu}}: MPSI), 10.7\,cm radio flux\break (\href{www.ngdc.noaa.gov/stp/space-weather/solar-data/solar-features/solar-
radio/noontime-flux/penticton/}{\textsf{ngdc.noaa.gov}}: F$_{10.7}$; \opencite{1987JGR....92..829T}), Physikalisch-Meteorologisches Observatorium Davos (PMOD) composite of Total Solar Irradiance (\href{ftp://ftp.pmodwrc.ch/pub/data/irradiance/composite/}{\textsf{ftp.pmodwrc.ch}}: TSI: \opencite{2006SSRv..125...53F}), Wilcox Solar Observatory Solar Mean Magnetic Field (\href{http://wso.stanford.edu/\#MeanField}{\textsf{wso.stanford.edu}}:\break SMMF: \opencite{1977SoPh...54..353S}), and Wilcox Solar Observatory Polar Fields \linebreak (\href{http://wso.stanford.edu/\#MeanField}{\textsf{wso.stanford.edu}}: \opencite{1978SoPh...58..225S}).  F$_{10.7}$ is expressed in Radio Flux Units [RFU: $\rm1\,RFU=10^{-22}\,W\,m^{-2}\,Hz^{-1}$]. In addition to considering the SSA data for the whole disk we also consider the SSA data for the northern (SSAN) and southern (SSAS) hemispheres separately. 

\section{Activity Measures of Solar Minimum}\label{section[proxies]}
\begin{table*}\caption{Activity proxies during the two solar minima. $N_0$ is the number of zero values observed during the respective minima.}\label{table[proxies]}
\setlength{\tabcolsep}{5pt}
\resizebox{\textwidth}{!}{\begin{tabular}{ccccccc}
  \hline
 Proxy & \multicolumn{3}{c}{$\rm Min_{23}$} & \multicolumn{3}{c}{$\rm Min_{24}$} \\
  & Mean & Median & $N_0$ & Mean & Median & $N_0$ \\
  \hline
  SSN & $11\pm11$ & $9\pm9$ & 269 & $3\pm6$ & $0\pm0$ & 588 \\
  SSA [msh] & $91\pm162$ & $28\pm28$ & 334 & $32\pm92$ & $0\pm0$ & 589 \\
  SSAN [msh] & $42\pm79$ & $0\pm0$ & 487 & $5\pm18$ & $0\pm0$ & 741 \\
  SSAS [msh] & $49\pm136$ & $0\pm0$ & 565 & $26.8\pm89$ & $0\pm0$ & 659\\
  MWSI & $0.02\pm0.05$ & $0\pm0$ & 391 & $0.01\pm0.04$ & $0\pm0$ & 536\\
  MPSI & $0.15\pm0.13$ & $0.12\pm0.08$ & 1 & $0.07\pm0.09$ & $0.03\pm0.02$ & 0 \\
  $\rm F_{10.7}$ [RFU] & $73\pm5$ & $72\pm2$ & n/a & $70\pm3$ & $69\pm1$ & n/a \\   
  TSI $\rm [W\,m^{-2}] $ & $1360.72\pm0.13$ & $1360.72\pm0.08$ & n/a & $1360.51\pm0.08$ & $1360.49\pm0.05$ & n/a \\
  SMMF [G] & $0.10\pm0.09$ & $0.08\pm0.05$ & n/a & $0.08\pm0.07$ & $0.05\pm0.03$ & n/a \\
  Polar field N [G] & $0.98\pm0.02$ & $0.99\pm0.01$ & n/a & $-0.52\pm0.04$ & $-0.54\pm0.03$ & n/a\\
  Polar field S [G] & $-0.90\pm0.10$ & $-0.86\pm0.04$ & n/a & $0.54\pm0.02$ & $0.53\pm0.02$ & n/a \\
  \hline
\end{tabular}}
\end{table*}

\begin{figure}
  \centering
  \includegraphics[clip, width=0.32\textwidth]{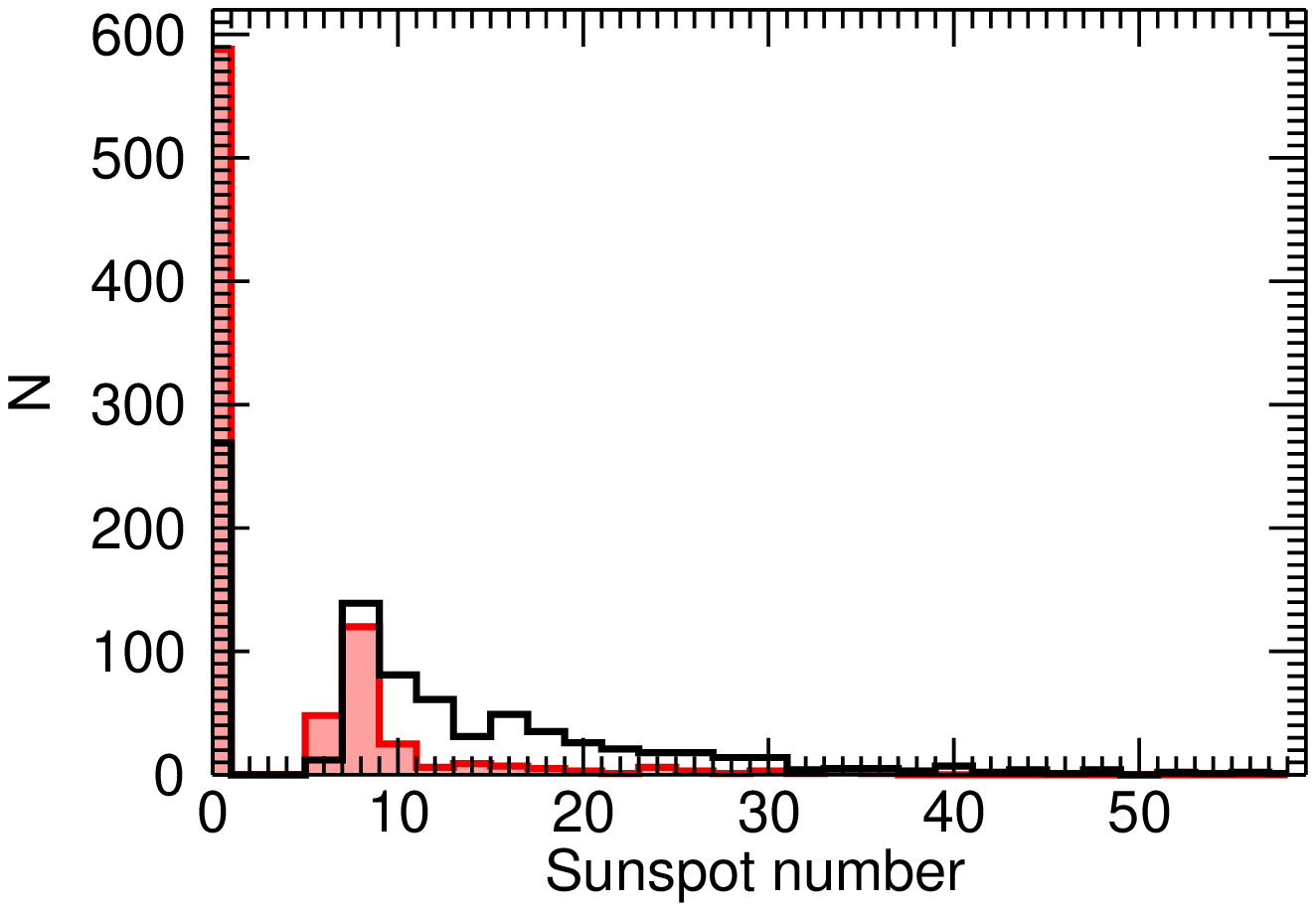}
   \includegraphics[clip, width=0.32\textwidth]{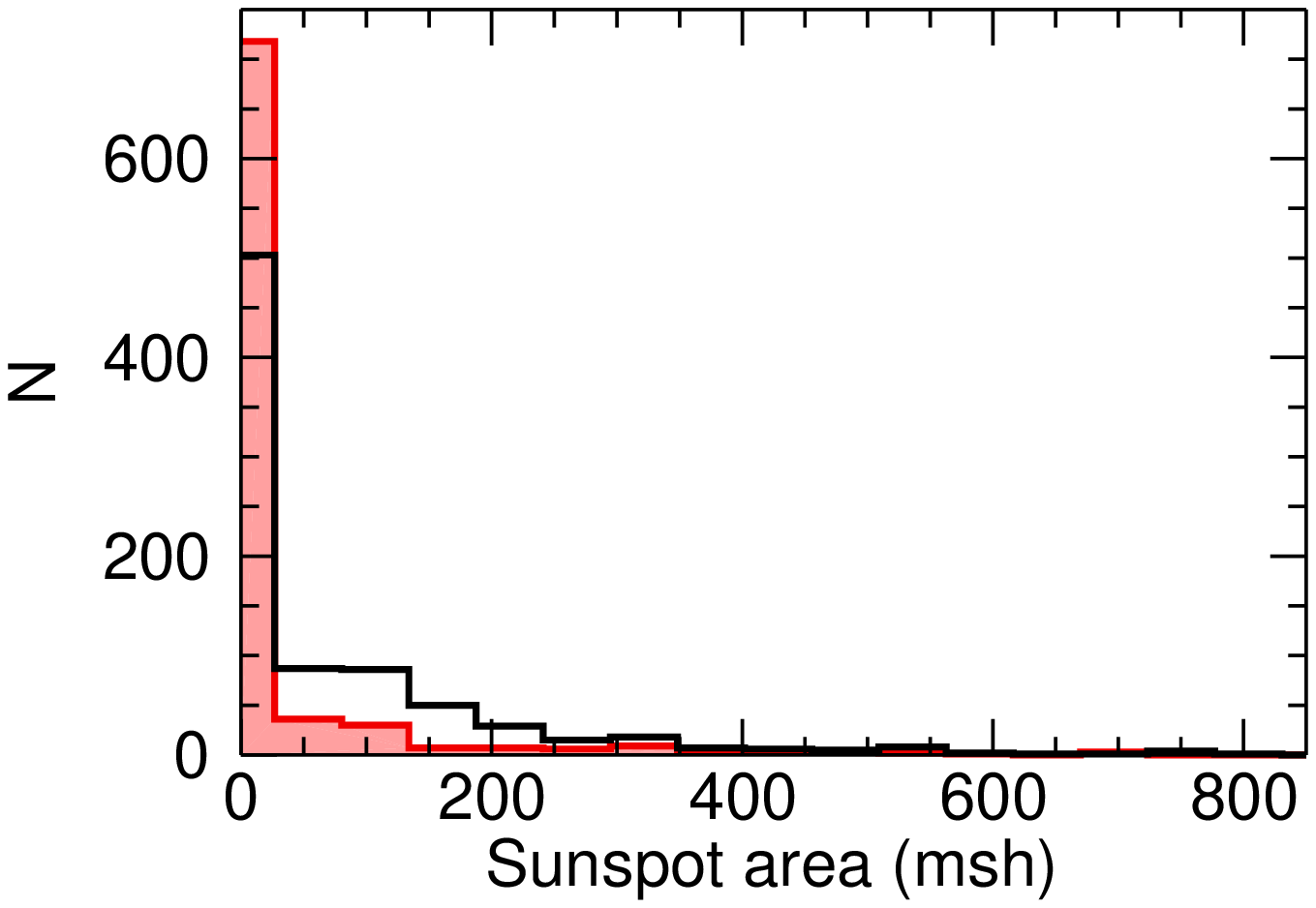}
    \includegraphics[clip, width=0.32\textwidth]{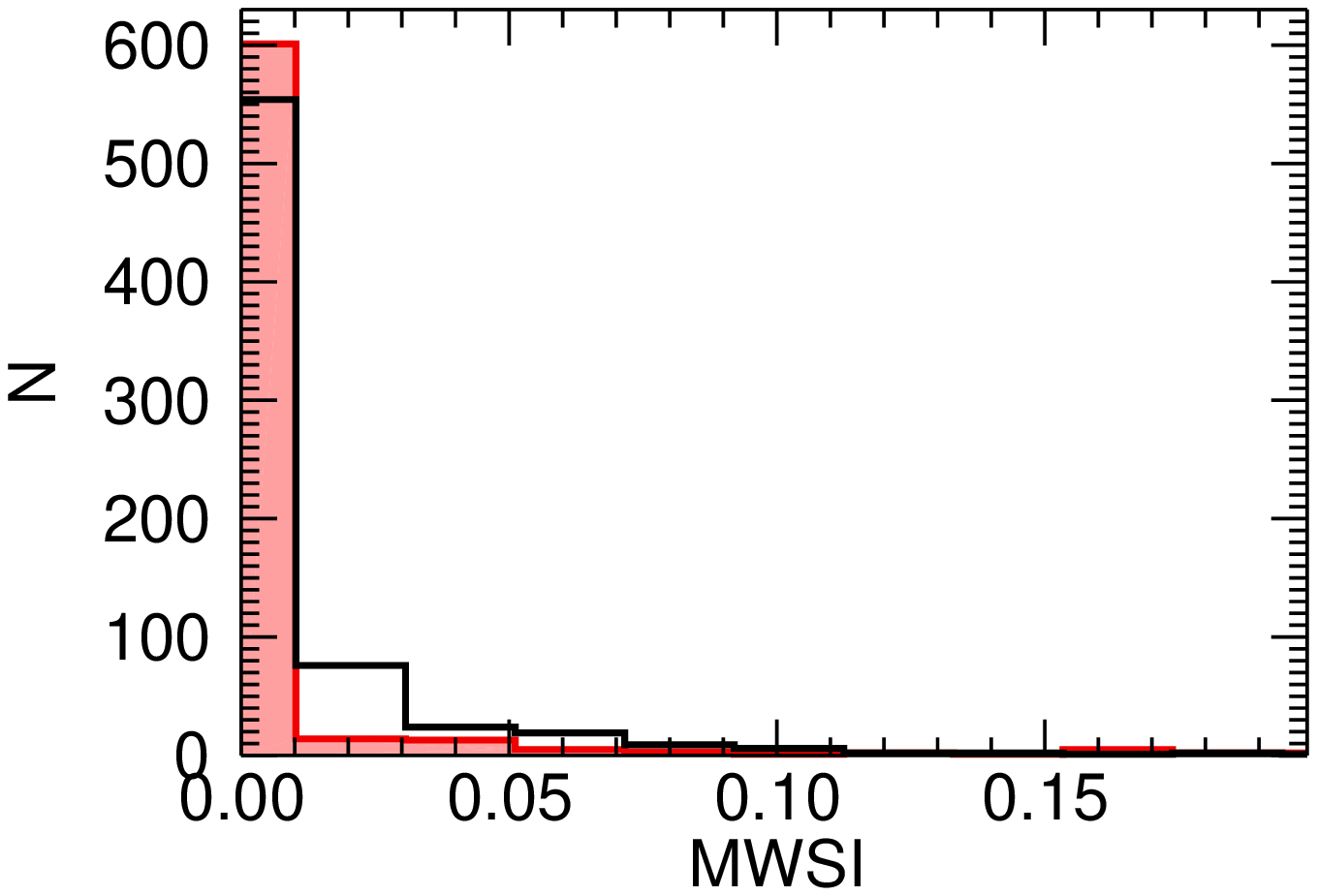}\\
  \includegraphics[clip, width=0.32\textwidth]{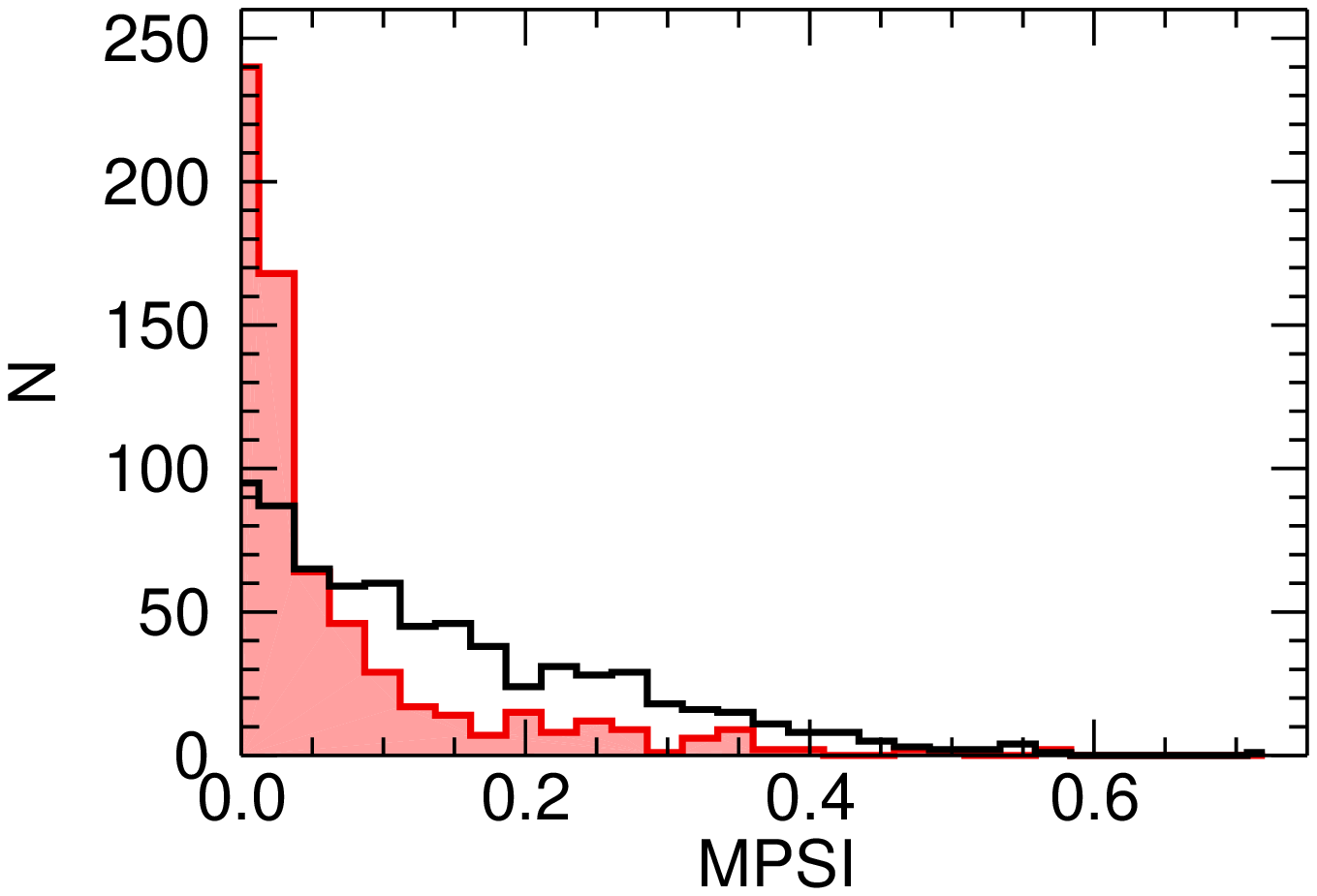}
   \includegraphics[clip, width=0.32\textwidth]{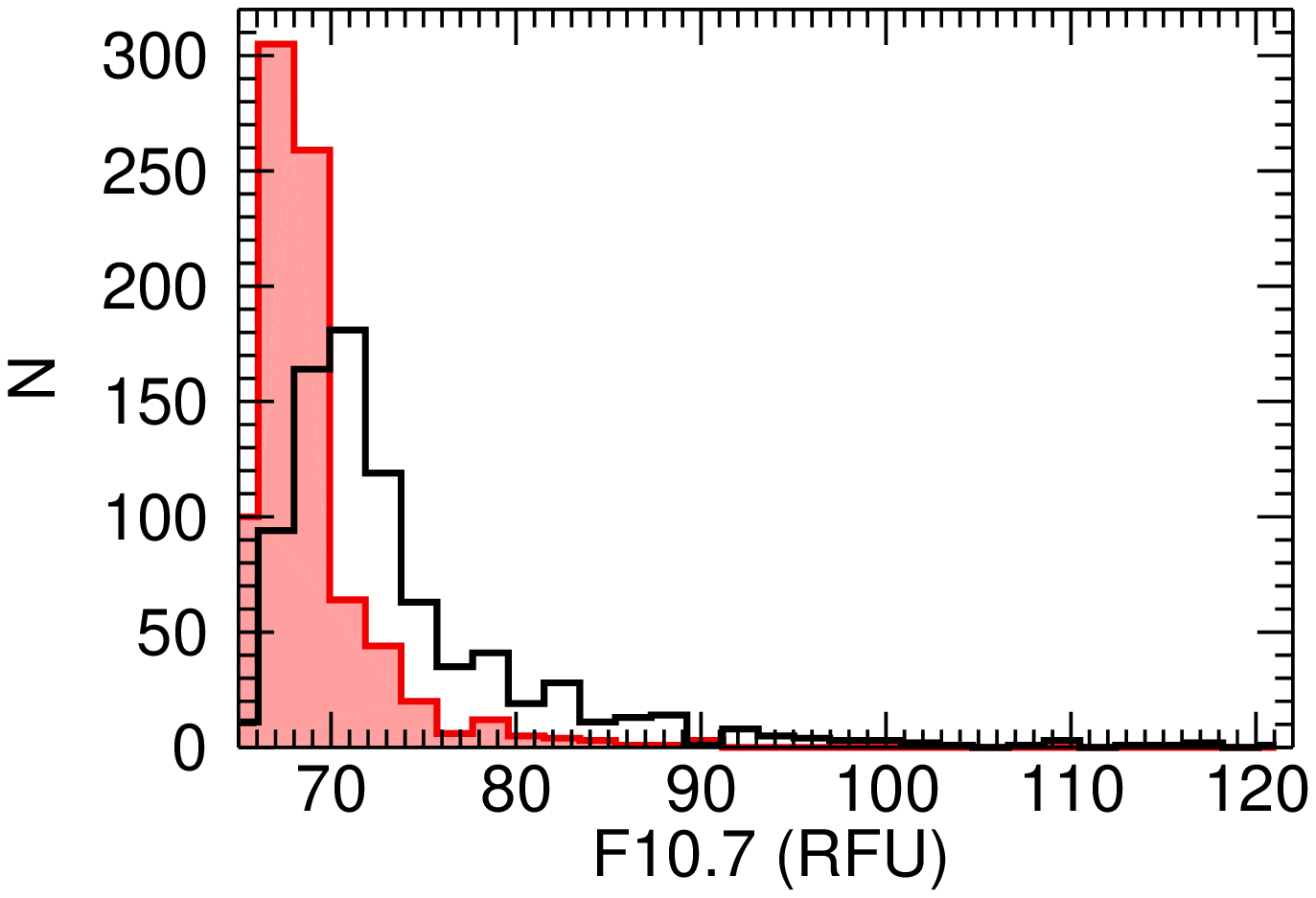}
    \includegraphics[clip, width=0.32\textwidth]{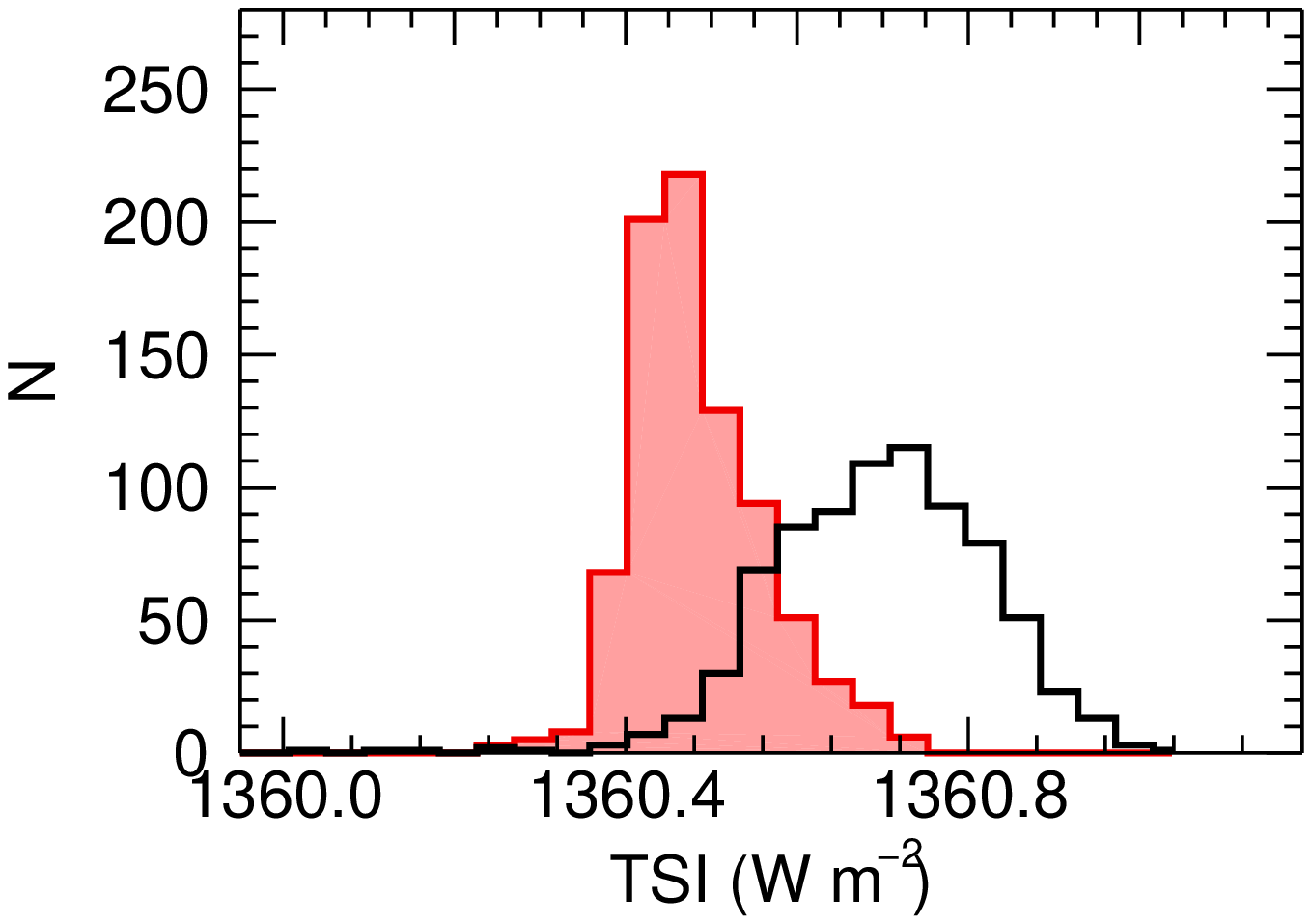}\\
     \includegraphics[clip, width=0.32\textwidth]{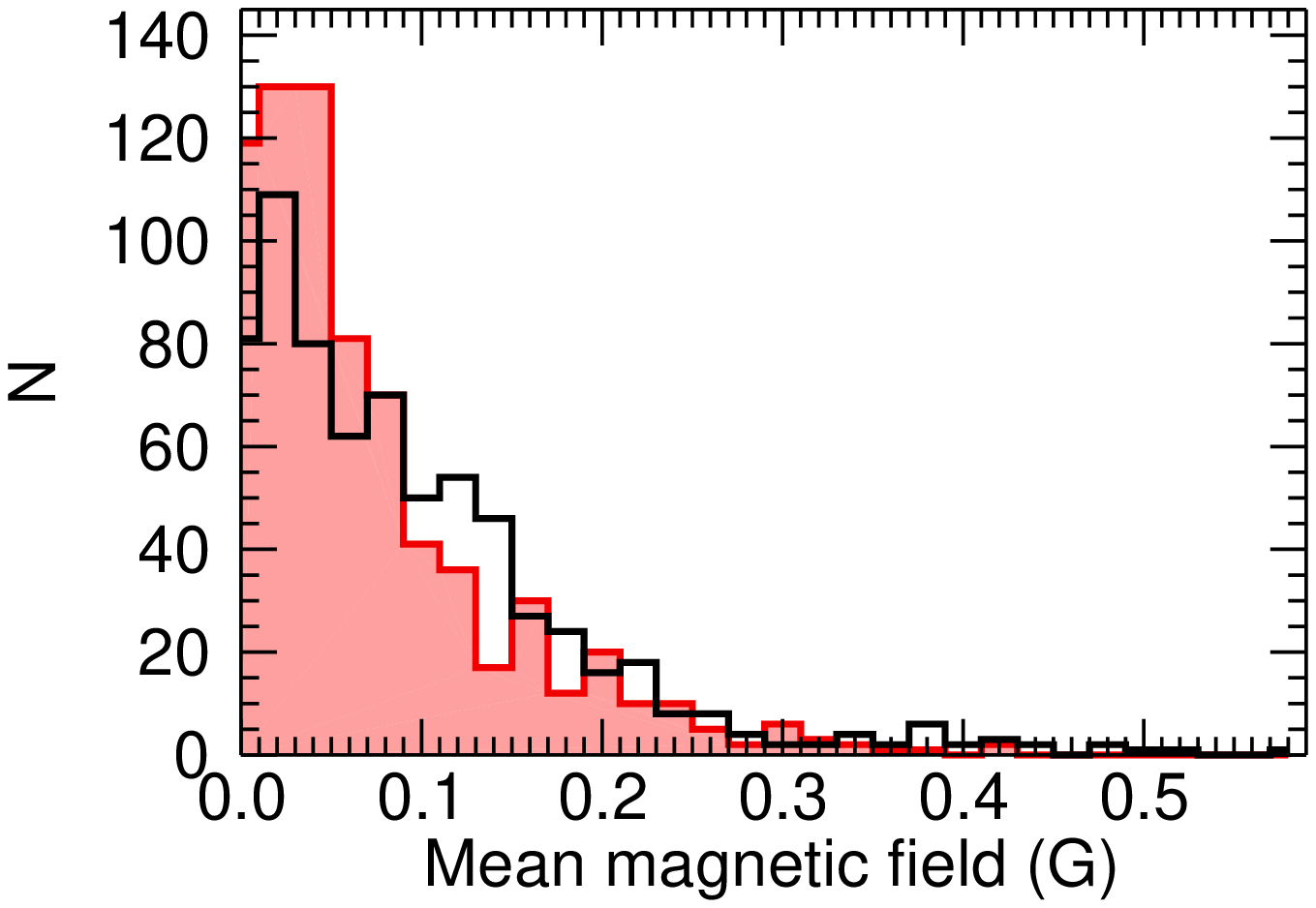}
      \includegraphics[clip, width=0.32\textwidth]{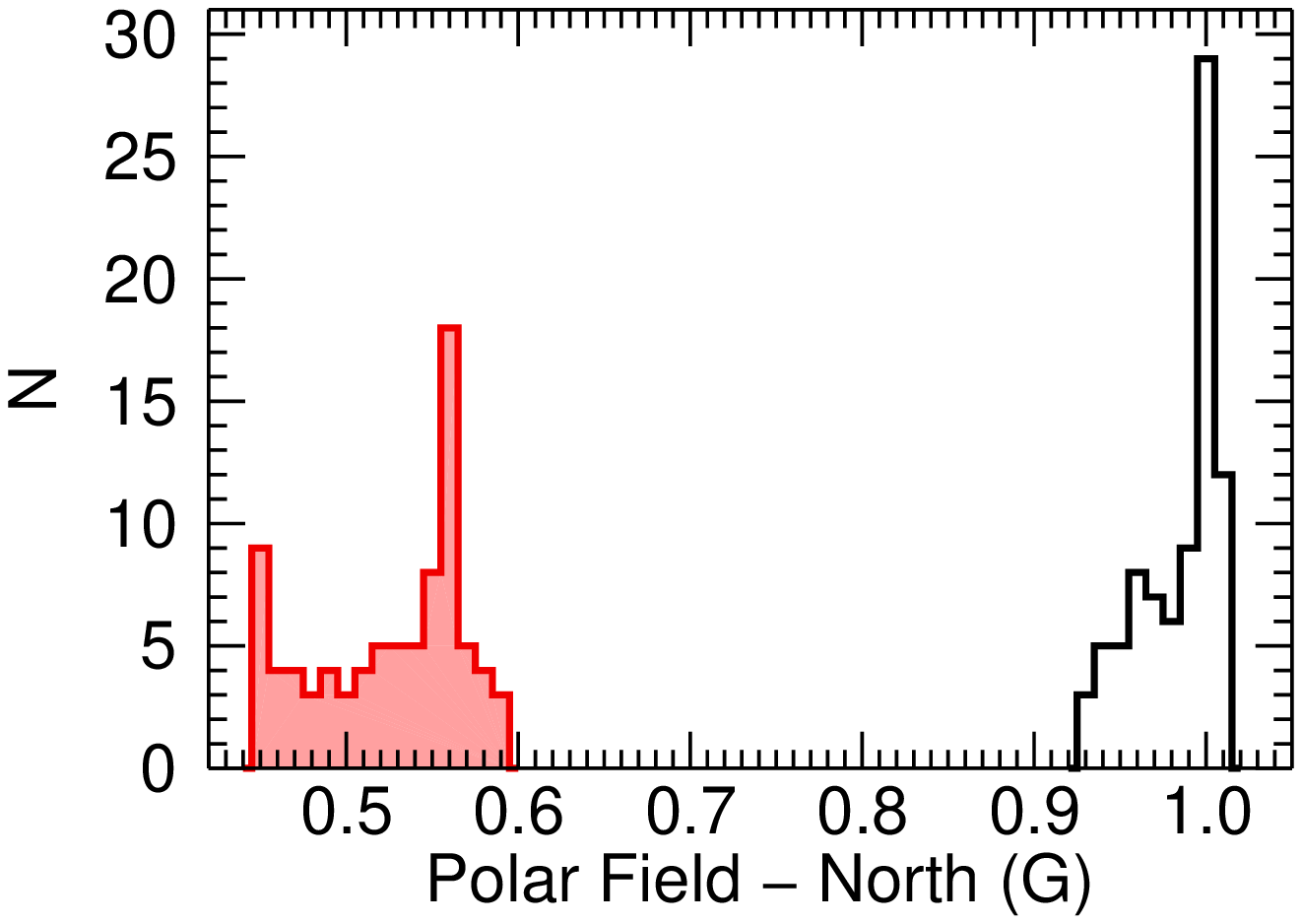}
       \includegraphics[clip, width=0.32\textwidth]{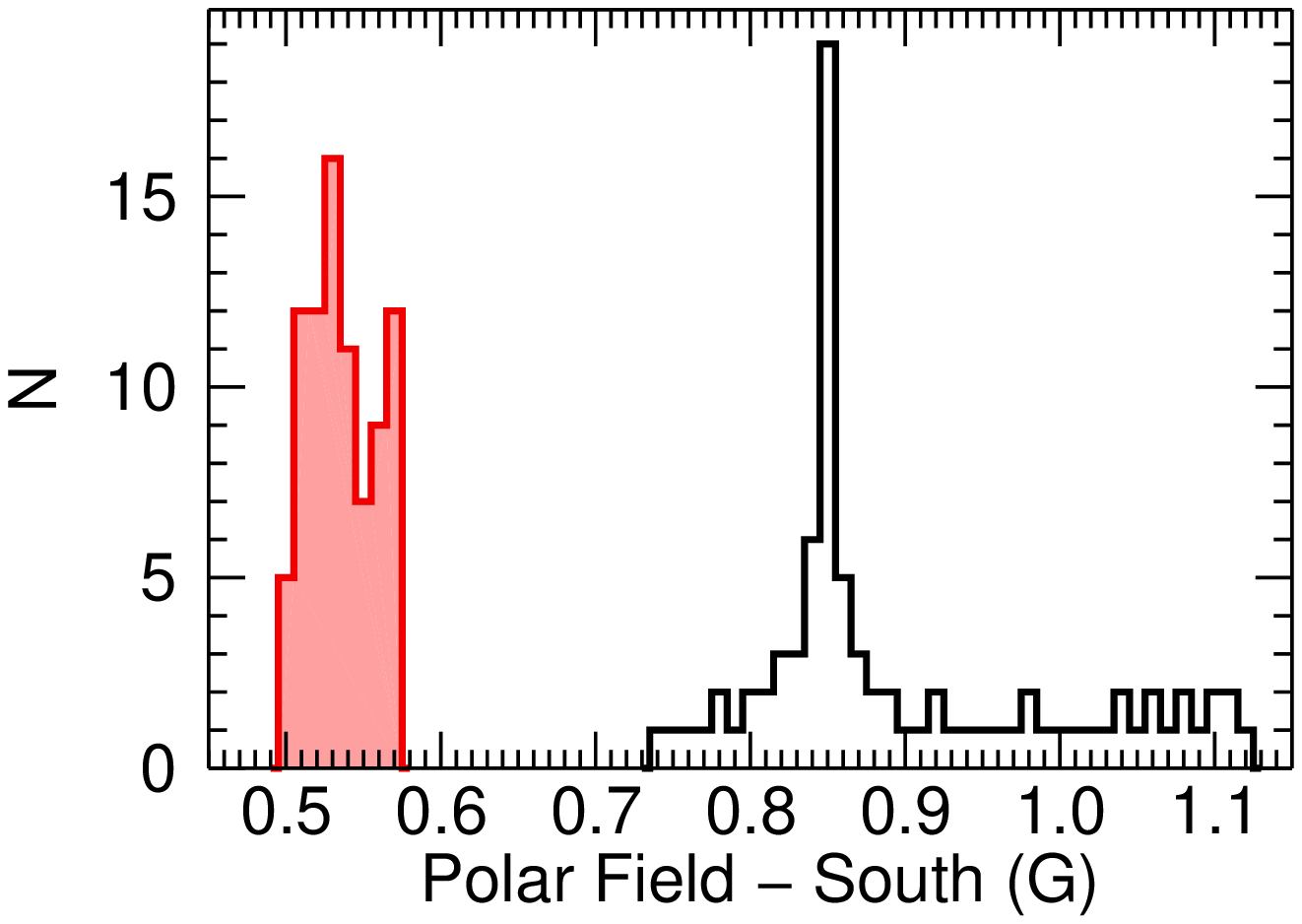}
  \caption{Histogram distribution of various solar-activity proxies observed during the two minima. The black lines with unshaded areas show the histograms for $\rm Min_{23}$, while the red lines with the shaded areas indicate the histograms for $\rm Min_{24}$. Top row (left to right): Sunspot number, sunspot area, MWSI. Middle row (left to right): MPSI, 10.7\,cm flux, TSI. Bottom row (left to right): Mean magnetic field, north polar field, south polar field. Note that the absolute values of the polar fields have been taken.}
  \label{figure[histograms]}
\end{figure}

Table \ref{table[proxies]} contains the average values of different activity measures for the last two activity minima. Figure \ref{figure[histograms]} shows the distributions of these activity proxies during the two minima.  The distributions of a number of the included activity measures are heavily skewed because they are zero-limited, specifically those related to sunspots (SSN, SSA, and MWSI) and MPSI. Therefore, the mean values quoted in Table \ref{table[proxies]} are not appropriate measures of the observed differences and are only included for completeness. However, we do note that for all sunspot measures considered here substantially more spotless days (\textit{i.e.} when SSN, SSA, MWSI were zero) were observed in $\rm Min_{24}$ than in $\rm Min_{23}$. The top row of Figure \ref{figure[histograms]} highlights the fact that the sunspot proxies are dominated by zero values in both minima, although the larger number of non-zero values in $\rm Min_{23}$ is visible. MPSI shows a similarly skewed histogram but, again, the tendency towards more higher values in $\rm Min_{23}$ can also be observed. On average the difference in the two minima as seen by F$_{10.7}$ is insignificant (see Table \ref{table[proxies]}). Nevertheless, the histograms do appear to indicate a semi-hard lower limit: rather than showing the same shape distributions but offset, the distribution for $\rm Min_{24}$ is restricted to a smaller range of values than $\rm Min_{23}$ but the lower limit is the same. The difference in the mean values of the TSI between the two minima is less than $2\sigma$.  However, the difference in the distributions of the observed TSI for the two minima is clear. The difference in the mean value of the SMMF is less than $1\sigma$ and this is also reflected in the distributions, which are similar. Conversely, both the northern and southern polar magnetic fields were substantially weaker during  $\rm Min_{24}$, with clearly separated distributions. Despite the clear difference in the polar magnetic fields, the case for the most recent minimum being significantly deeper than the previous one is not clear cut, with the majority of the activity proxies examined here showing significant overlap. We now show that the helioseismic frequencies are a robust measure of the depth of the solar minimum.

\section{Helioseismic Comparison of Minima}\label{section[results]}

\begin{figure}
  \centering
  \includegraphics[clip, width=0.35\textwidth, angle=90]{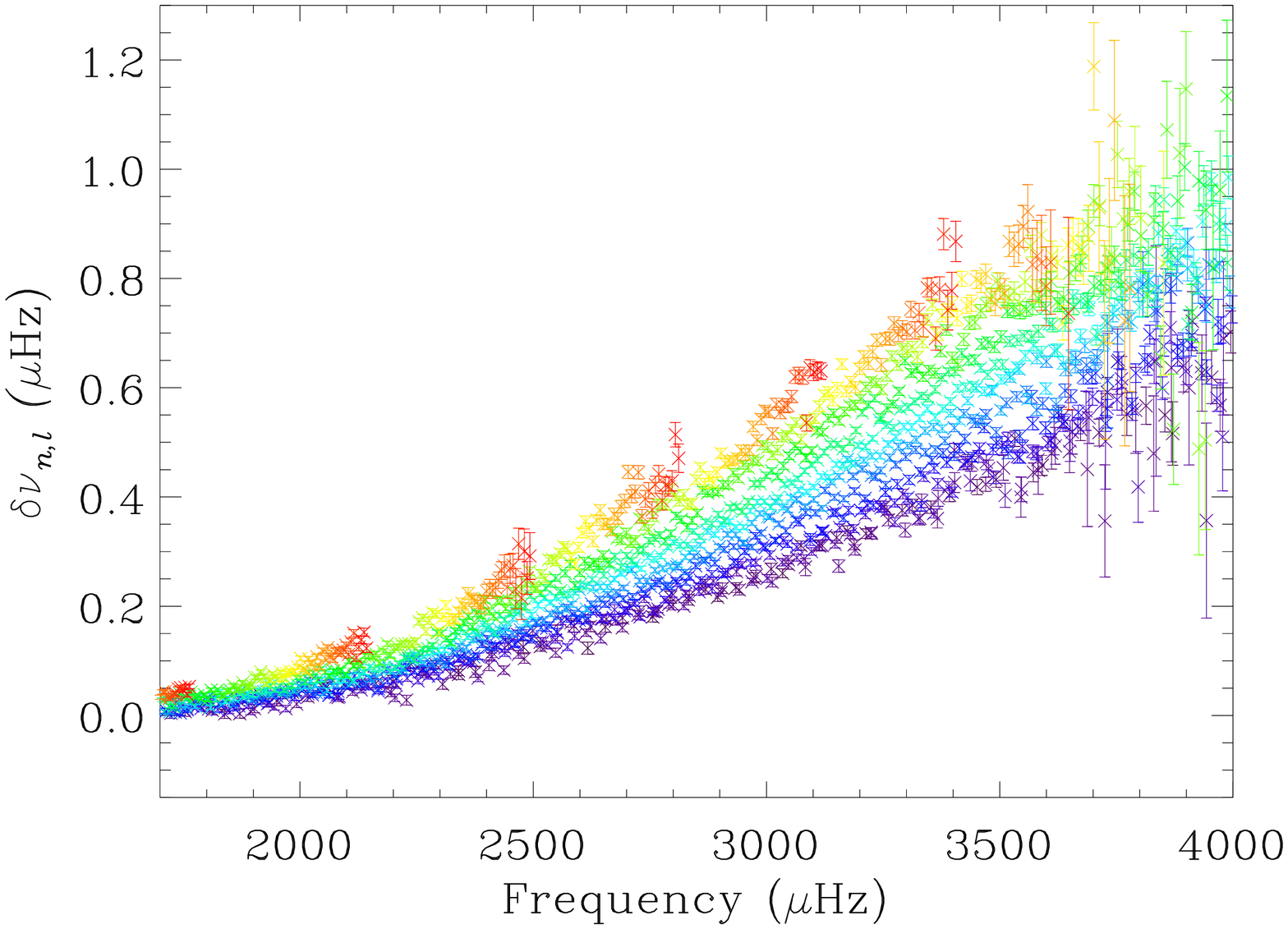} 
  \includegraphics[clip, width=0.35\textwidth, angle=90]{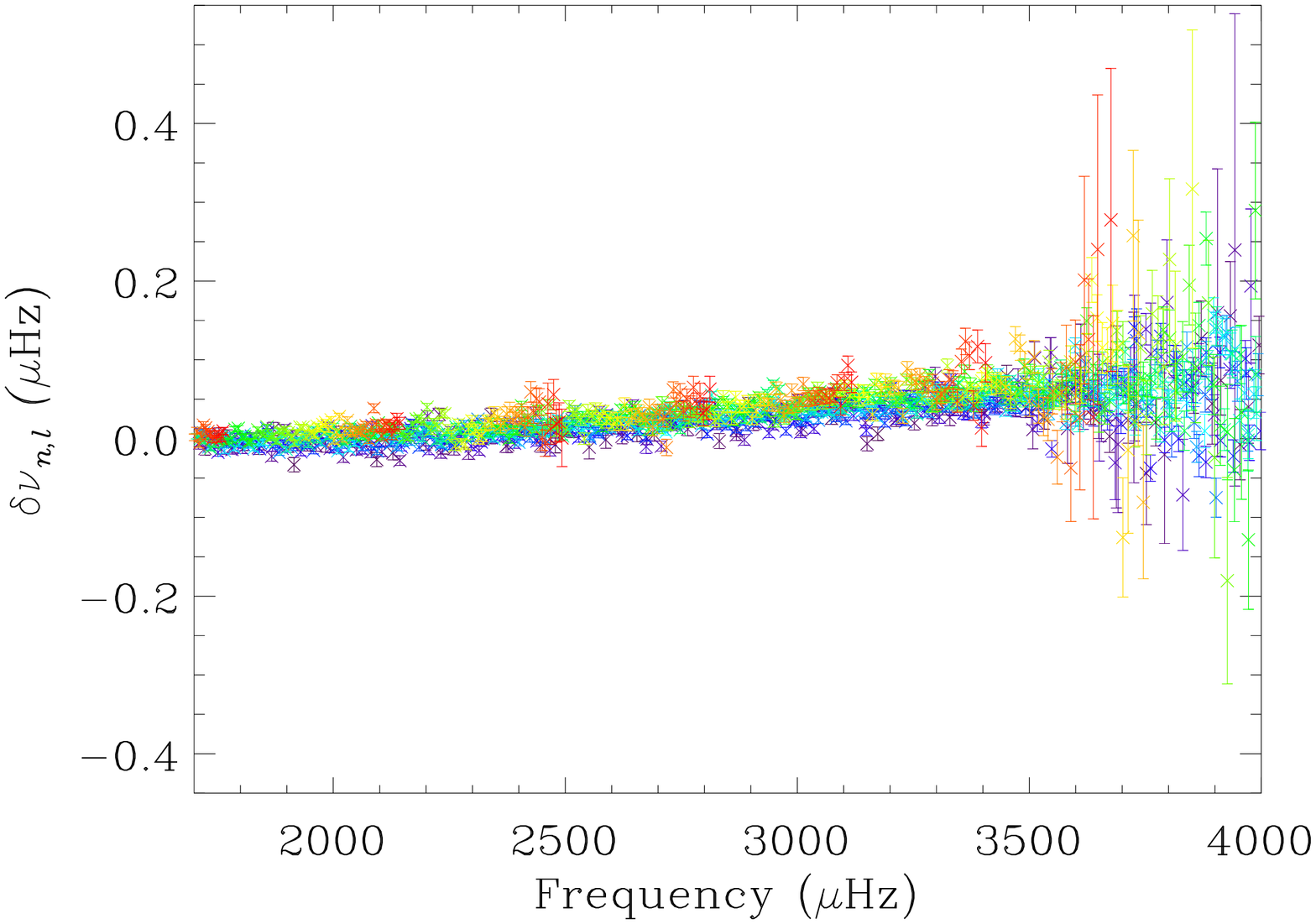}\\
  \includegraphics[clip, width=0.35\textwidth, angle=90]{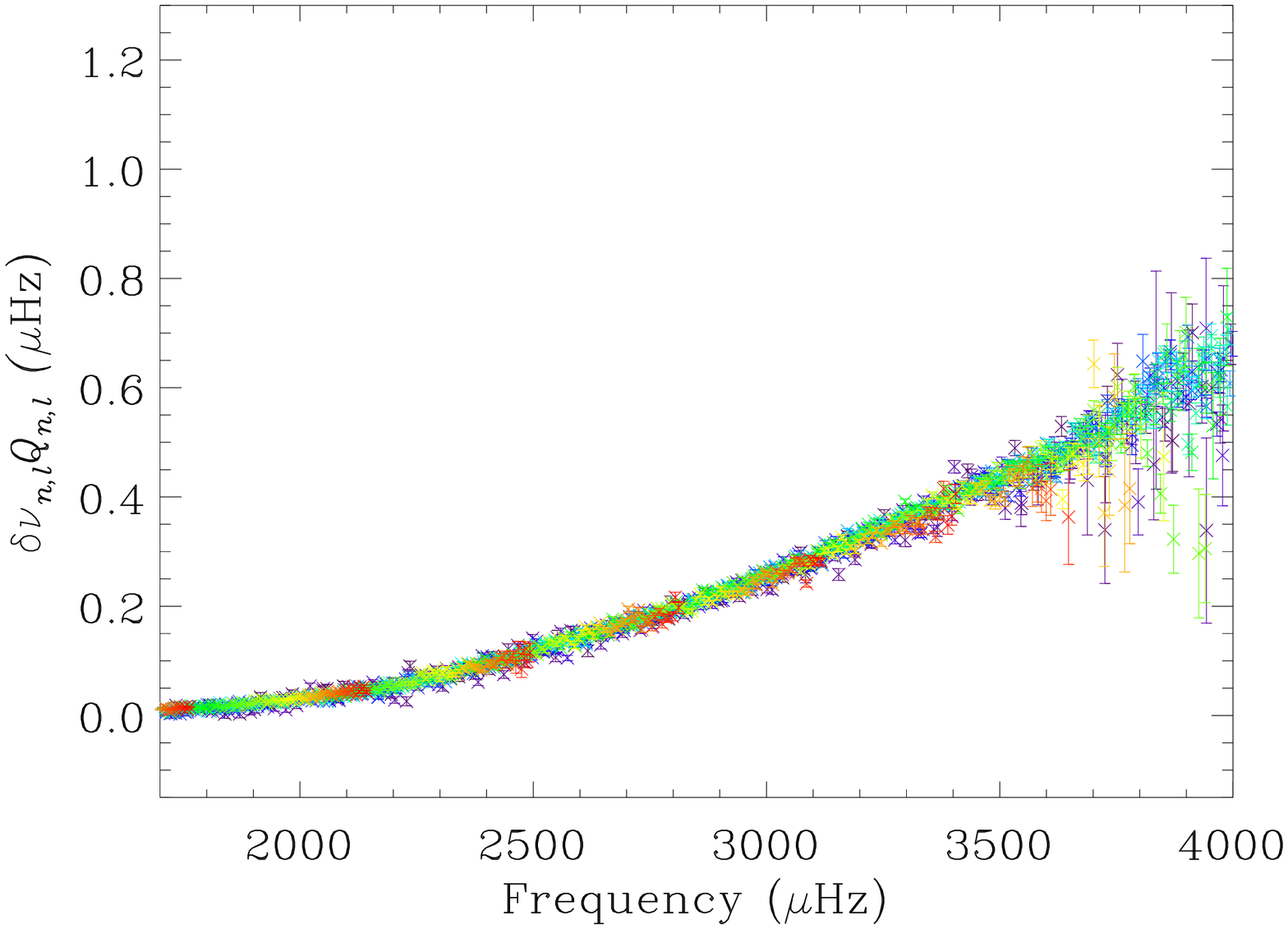}
  \includegraphics[clip, width=0.35\textwidth, angle=90]{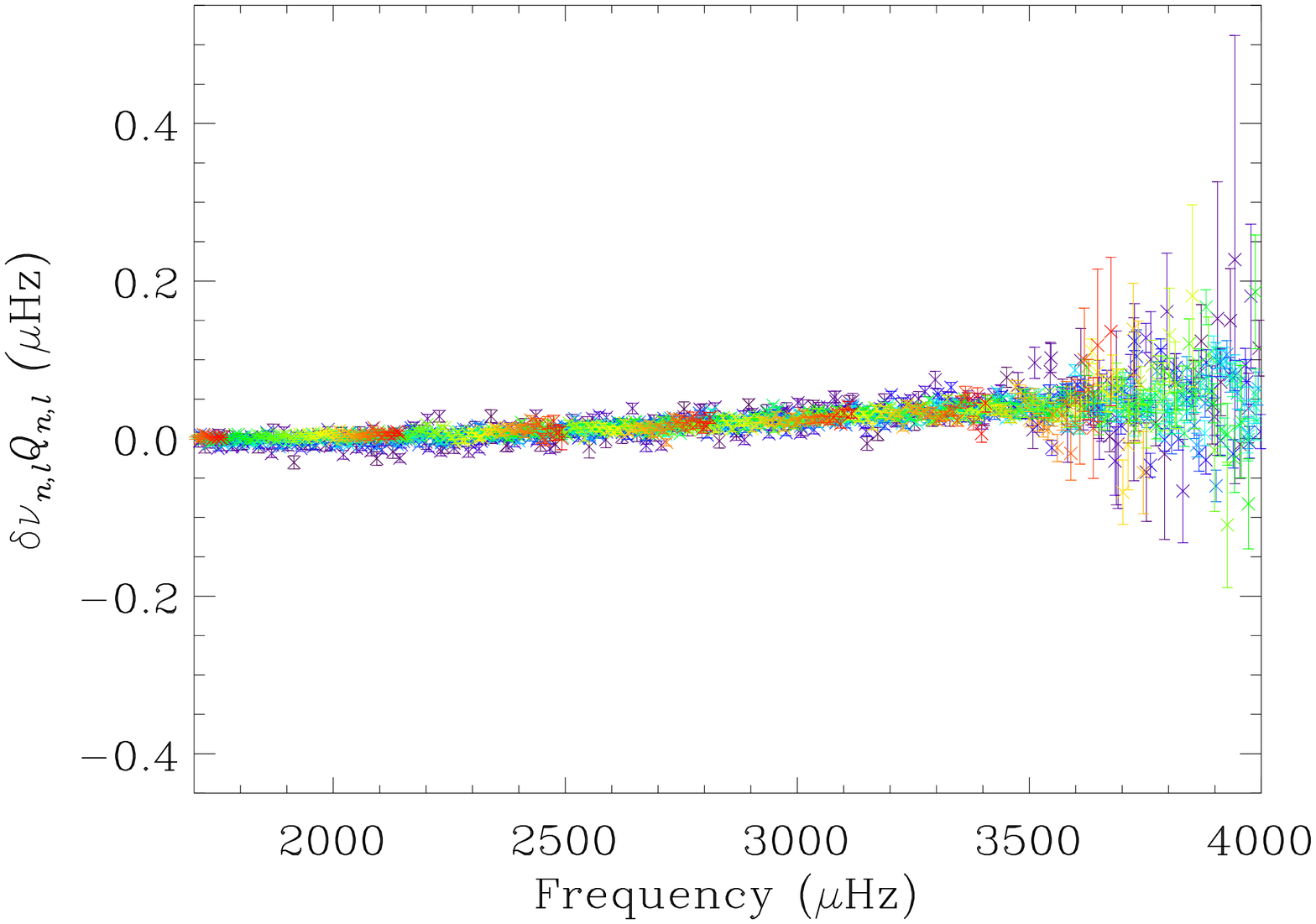}\\
  \includegraphics[clip, width=0.35\textwidth, angle=90]{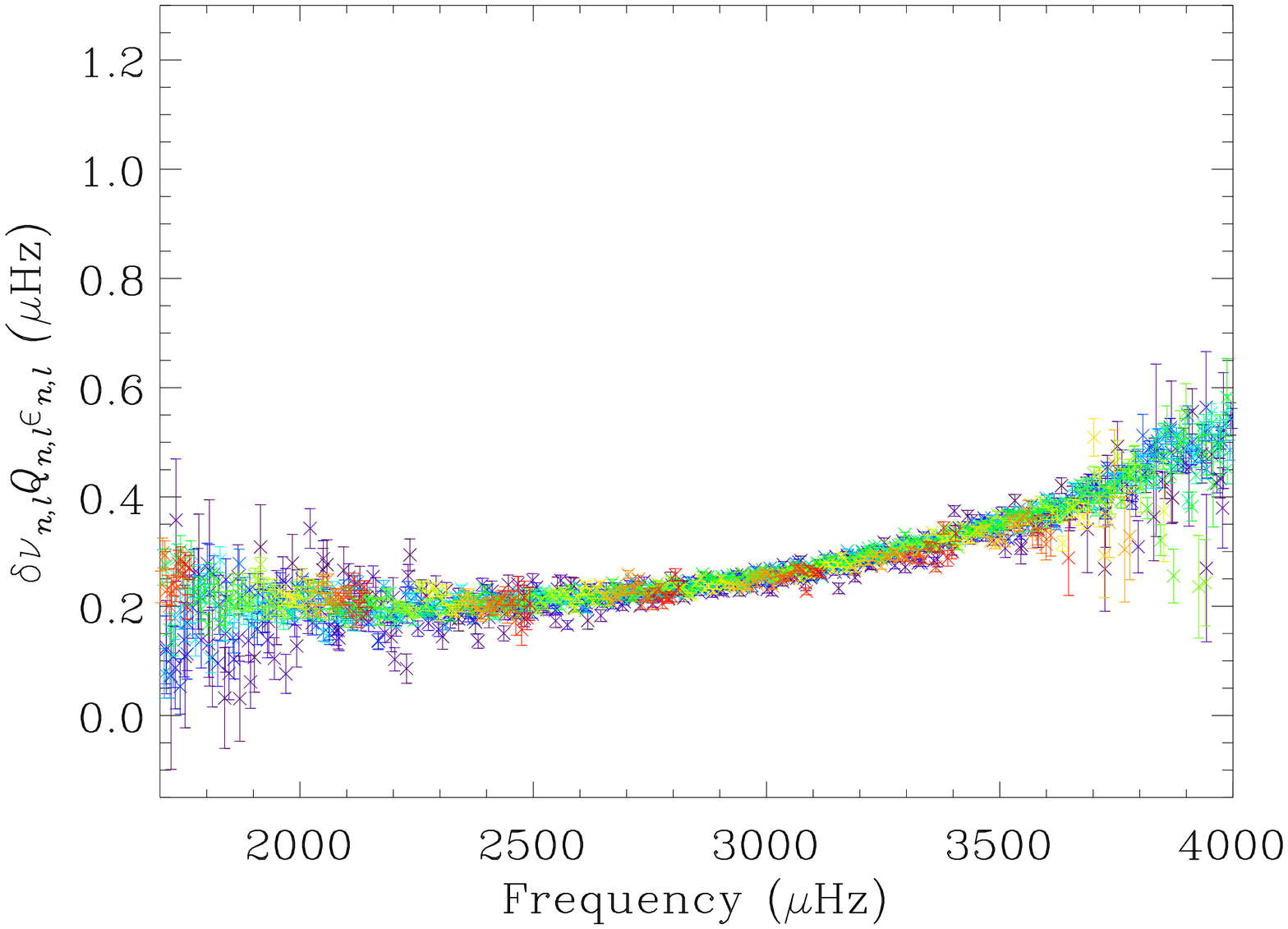}
  \includegraphics[clip, width=0.35\textwidth, angle=90]{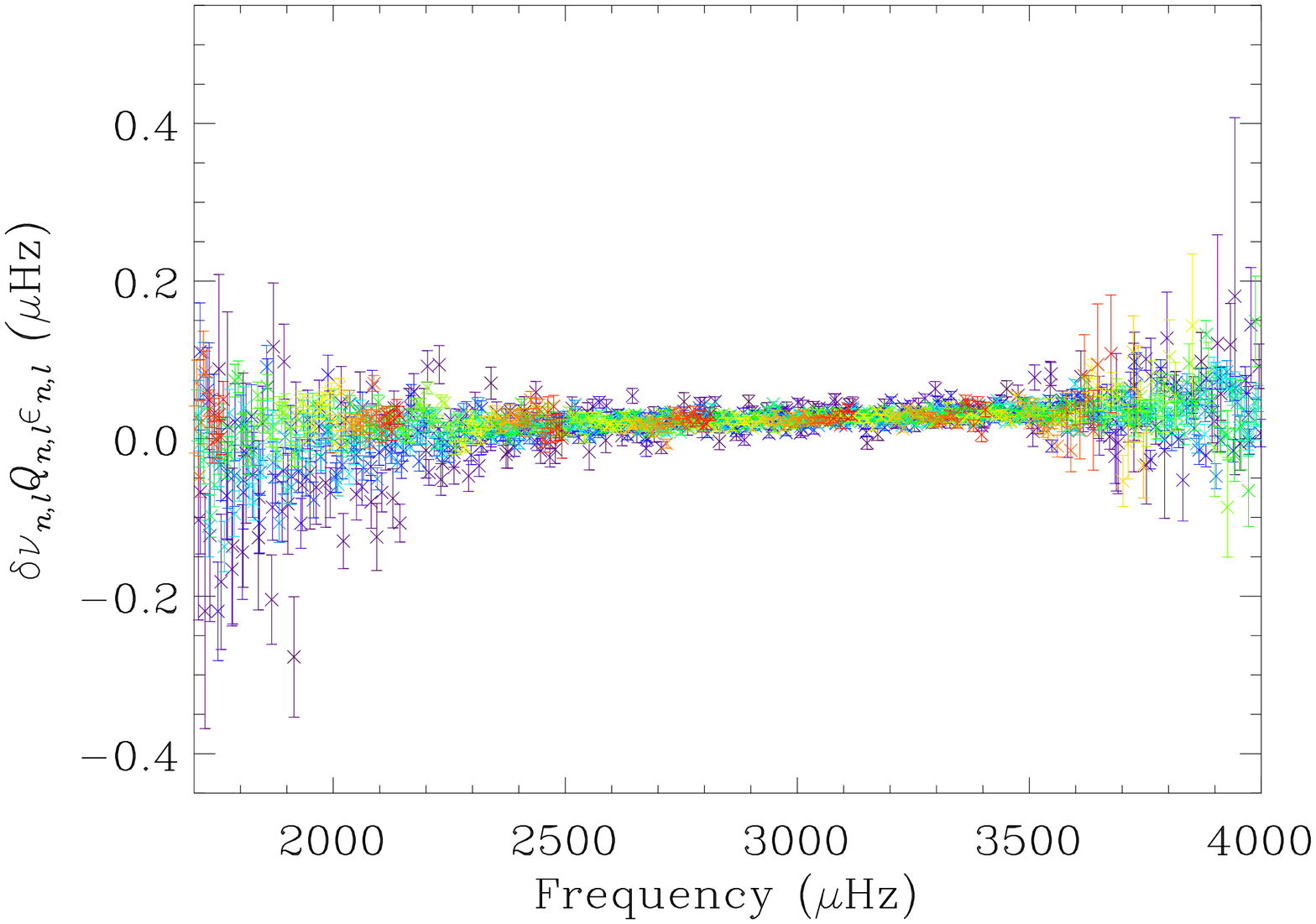}\\
  \includegraphics[clip, width=0.5\textwidth]{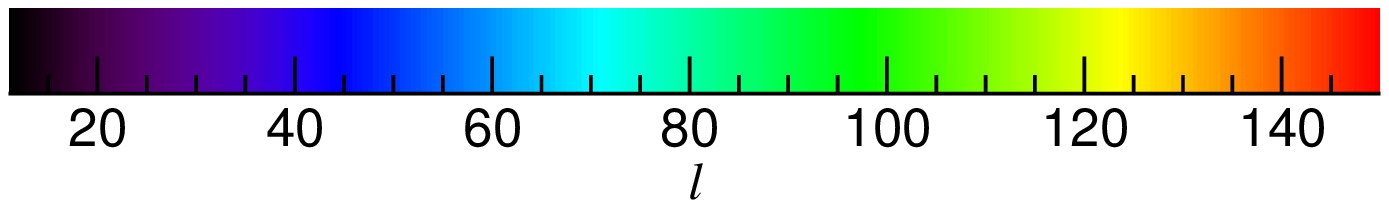}
  \caption{Top-left: Raw frequency shift [$\delta\nu_{n,\ell}$] observed between $\rm Max_{23}$ and $\rm Min_{23}$. Top-right: Raw frequency shift [$\delta\nu_{n,\ell}$] observed between $\rm Min_{23}$ and $\rm Min_{24}$. Middle-left: Inertia-corrected frequency shift [$\delta\nu_{n,\ell}Q_{n,\ell}$] observed between $\rm Max_{23}$ and $\rm Min_{23}$. Middle-right:  Inertia-corrected frequency shift [$\delta\nu_{n,\ell}Q_{n,\ell}$] observed between $\rm Min_{23}$ and $\rm Min_{24}$.  Bottom-left: Inertia-corrected and inertia-frequency-corrected frequency shift [$\delta\nu_{n,\ell}Q_{n,\ell}\epsilon_{n,\ell}$] observed between $\rm Max_{23}$ and $\rm Min_{23}$. Bottom-right:  Inertia-corrected and inertia-frequency-corrected frequency shift [$\delta\nu_{n,\ell}Q_{n,\ell}\epsilon_{n,\ell}$] observed between $\rm Min_{23}$ and $\rm Min_{24}$. In all plots colours represent different harmonic degrees [$\ell$], as indicated by the colourbar.}
  \label{figure[frequency_dependence]}
\end{figure}

The weighted mean $m$-averaged frequency of each mode was determined at each of the two minima and the activity maximum (as defined in Section \ref{section[data]}), where the weights were the inverse square of the formal uncertainties associated with the frequencies. The top two panels of Figure \ref{figure[frequency_dependence]} show the observed change in frequency [$\delta\nu$] as a function of $m$-averaged frequency. As expected the shift between cycle maximum and cycle minimum (top-left hand panel of Figure \ref{figure[frequency_dependence]}) is dependent on both mode frequency and $\ell$ \citep[see][for a recent review]{2014SSRv..186..191B}. Although the magnitude of the frequency shift is reduced, similar $\ell$ and frequency dependencies are observed for the frequency shifts between the two minima (top right-hand panel of Figure \ref{figure[frequency_dependence]}). 
 
To remove the dependence on $\ell$, the frequencies were corrected for mode inertia in the following manner \citep[see also][]{2001MNRAS.324..910C}. The mode inertia [$I_{n,\ell}$] is defined as the internal mass of the Sun affected by the oscillation [$M_{n,\ell}$] relative to the total mass of the Sun [$\textrm{M}_\odot$] and is given by \citep{1991sia..book..401C}
\begin{equation}\label{equation[mode inertia]}
   I_{n,l}=\textrm{M}_\odot^{-1}\int_{v}|\xi|^2\rho\textrm{d}V=
   4\pi \textrm{M}_{\odot}^{-1}\int_0^{\textrm{\scriptsize{R}}_\odot}|\xi|^2\rho r^2\textrm{d}r=
   \frac{M_{n,l}}{\textrm{M}_{\odot}},
\end{equation}
where $\xi$ is the photosphere-normalized mode displacement associated with a mode, $V$ is the volume of the Sun, $\rho$ is density, $r$ is distance in the radial direction and $R_\odot$ is the radius of the Sun. The inertia ratio [$Q_{n,\ell}$] is defined by \citet{1991sia..book..401C} as:
 \begin{equation}
 Q_{n\ell} = I_{n,\ell} / \bar{I}(\nu_{n, \ell}),
 \label{eqution[inertia ratio]}
 \end{equation}
$\bar{I}(\nu_{n,\ell})$ is the inertia that an $\ell=0$ mode would have at a frequency $\nu_{n, \ell}$. Values of $Q_{n,\ell}$ used in this study were derived from Model S \citep{1996Sci...272.1286C}. The middle two panels of Figure \ref{figure[frequency_dependence]} show the result of multiplying the frequency difference by $Q_{n,\ell}$ \textit{i.e.} $\delta\nu_{n,\ell}Q_{n,\ell}$. The degree dependence has been predominantly removed in both cases. 

Although primarily a function of degree, the mode inertia also depends on the frequency of the mode under consideration. It is, therefore, also possible to scale for this dependence. \citet{2001MNRAS.324..910C} define the ``inverse fractional mode inertia'' [$\epsilon^{-1}_{n,\ell}$] as 
\begin{equation}\label{equation[inverse mode inertia]}
  \epsilon^{-1}_{n,\ell}=\left(\frac{I_{n,\ell}}{\bar{I}(3000)}\right)^{-1}, 
\end{equation}
where $\bar{I}(3000)$ is the inertia that an $\ell=0$ mode would have at $3000\,\rm\mu Hz$. The bottom panels of Figure \ref{figure[frequency_dependence]} show how $\delta\nu_{n,\ell}Q_{n,\ell}\epsilon_{n,\ell}$ varies as a function of frequency.

To determine the significance of the frequency shifts observed between the two minima, average shifts over various frequency ranges were determined. First, the average frequency shift over the entire frequency range was calculated, and then the average frequency shifts observed in five different frequency bands. These values are shown in Table \ref{table[shifts]}. The frequency shifts are highly significant when the entire frequency range is considered and for each of the smaller frequency ranges, suggesting there is a small but significant offset in the oscillation frequencies between the two minima.

\begin{table}\caption{Observed average shifts in frequency between the two minima, where the average is taken over various different frequency bands. The shifts have been calculated with respect to $\rm Min_{24}$. }\label{table[shifts]}
\begin{tabular}{ccccccc}
  \hline
  Frequency range && $\delta\nu Q_{n,\ell}$  & N $\sigma$ &  & $\delta\nu Q_{n,\ell}\epsilon_{n, \ell}$  & N $\sigma$ \\
 $[\mu\rm Hz]$ && [$\rm nHz$] &  & & [$\rm nHz$] & \\ 
 \hline
  $1700\le\nu\le4000$ && $10.1\pm0.1$ & 114.9 & & $15.7\pm0.1$ & 178.8 \\
  $1700\le\nu\le2160$ && $1.3\pm0.1$ & 10.5 & & $12.3\pm1.0$ & 13.1 \\
  $2160\le\nu\le2620$ && $7.6\pm0.2$ & 38.7 & & $17.8\pm0.4$ & 41.5 \\
 $2620\le\nu\le3080$  && $20.3\pm0.2$ & 97.2 &  & $23.3\pm0.2$ & 98.9 \\
  $3080\le\nu\le3540$ && $33.6\pm0.3$ & 127.0 & & $29.2\pm0.2$ & 127.5 \\
 $3540\le\nu\le4000$ && $45.2\pm0.8$ & 59.4 & & $35.9\pm0.6$ & 59.4 \\
  \hline
\end{tabular}
\end{table}

\begin{figure}
  \centering
   \includegraphics[clip, width=0.35\textwidth, angle=90]{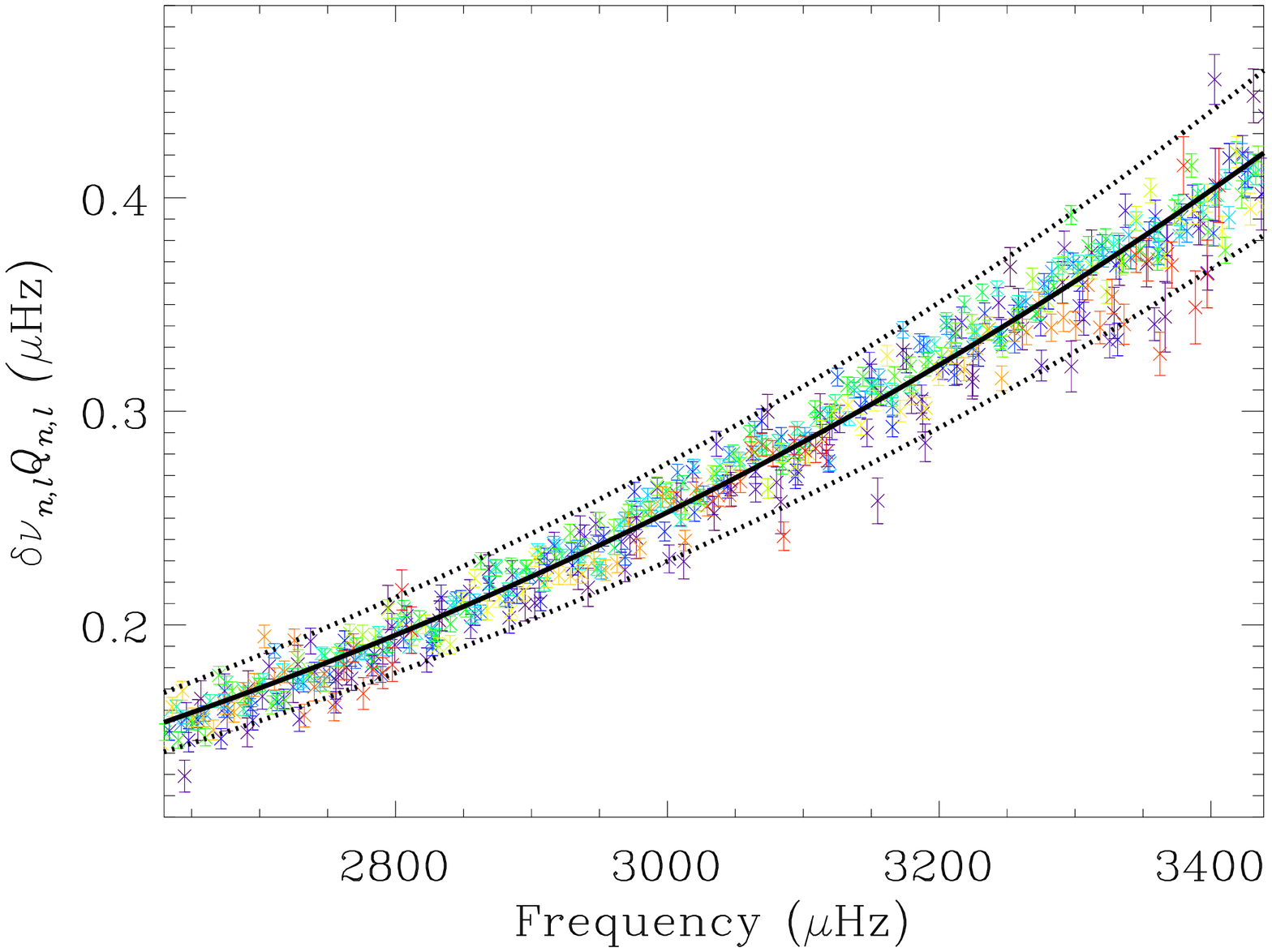}
  \includegraphics[clip, width=0.35\textwidth, angle=90]{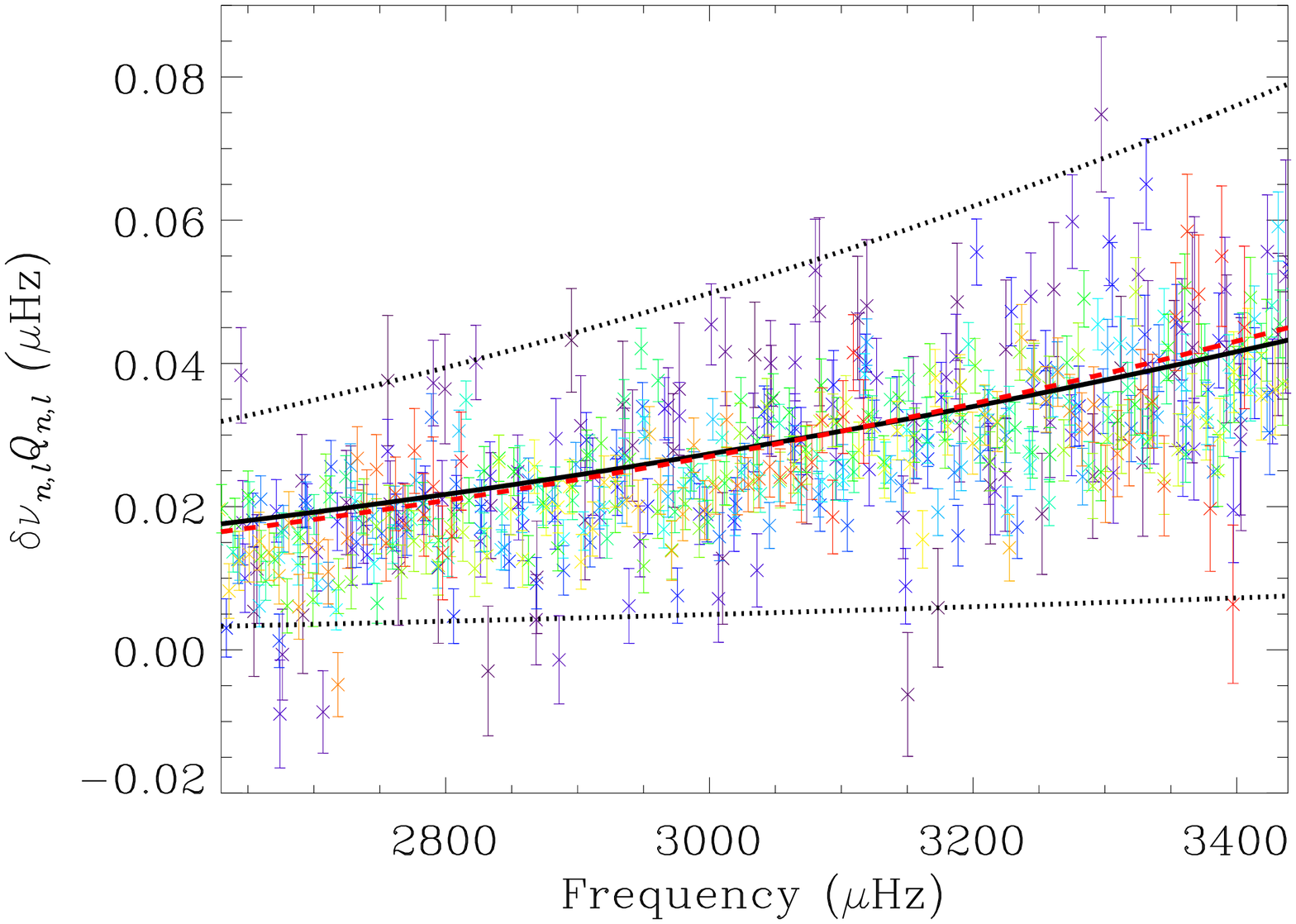}\\
  \includegraphics[clip, width=0.35\textwidth, angle=90]{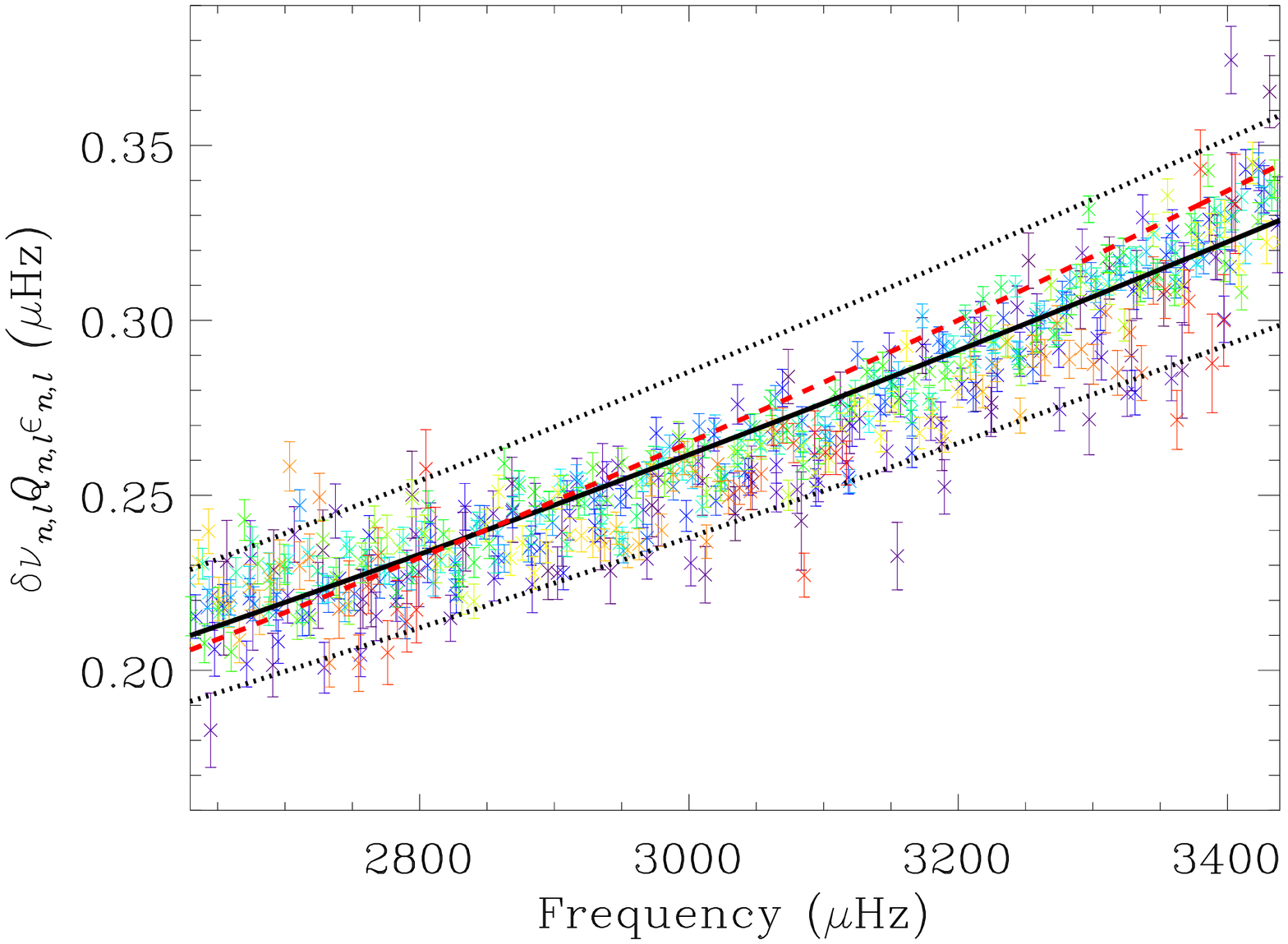}
  \includegraphics[clip, width=0.35\textwidth, angle=90]{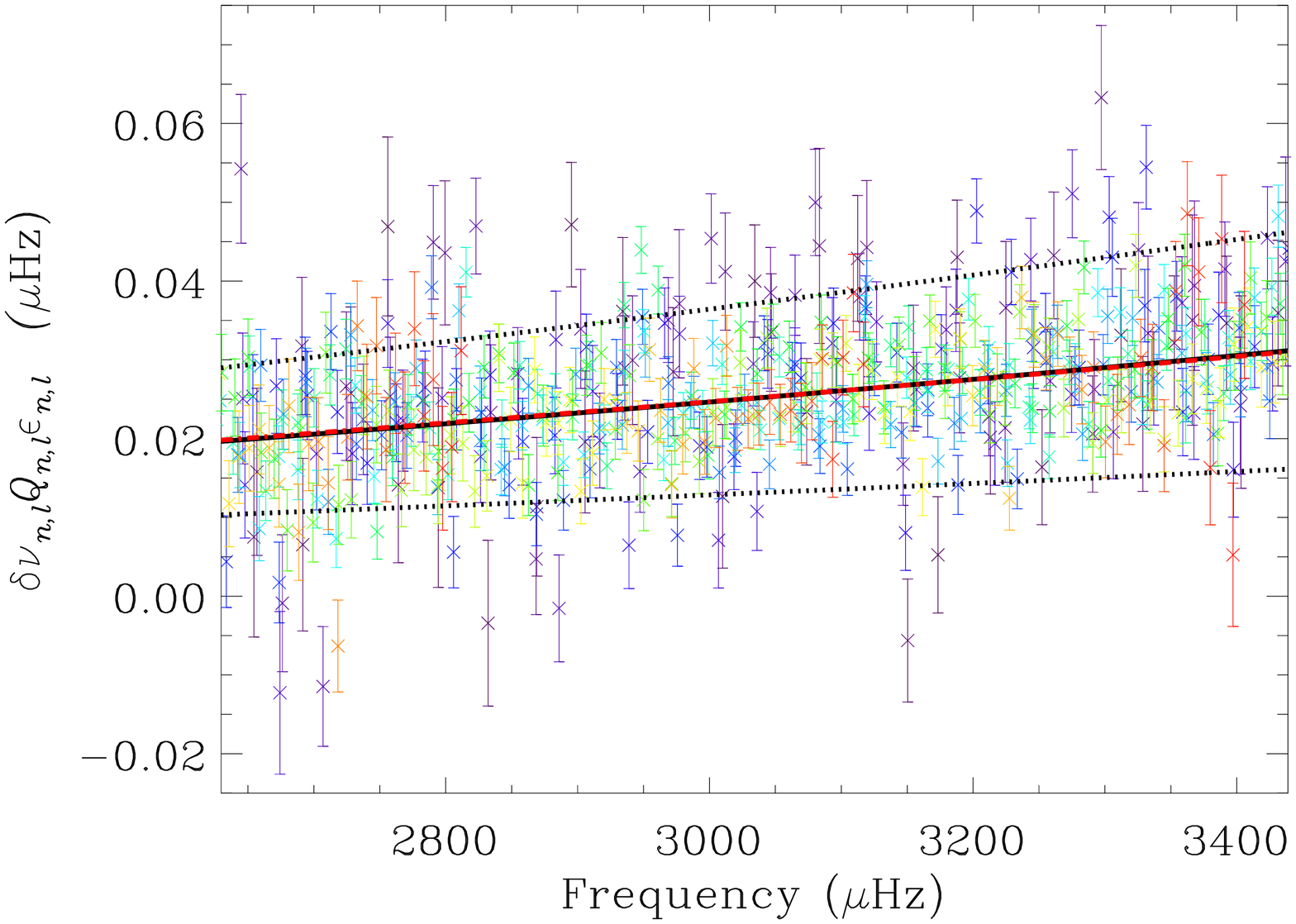}\\
  \includegraphics[clip, width=0.5\textwidth]{colorbar.eps}
  \caption{Top-left:  Inertia-corrected frequency shift $\left[\delta\nu Q_{n,\ell}\right]$ observed between $\rm Max_{23}$ and $\rm Min_{23}$. Overplotted in black is a fit to the data (solid line), described by Equation (\ref{eq[freq_shift]}), and associated uncertainties (dotted lines). Top-right: Inertia-corrected frequency shift $\left[\delta\nu Q_{n,\ell}\right]$ observed between $\rm Min_{23}$ and $\rm Min_{24}$. Overplotted in black is a fit to the data (solid line), described by Equation (\ref{eq[freq_shift]}), and associated uncertainties (dotted lines). Overplotted in red (dashed) is the fit where $\beta$ was obtained for the $\rm Max_{23}$--$\rm Min_{23}$ comparison. Bottom-left:  Inertia-corrected and inertia-frequency-corrected frequency shift $\left[\delta\nu Q_{n,\ell}\epsilon_{n,\ell}\right]$ observed between $\rm Max_{23}$ and $\rm Min_{23}$. Overplotted in black is a fit to the data (solid line), described by Equation (\ref{eq[freq_shift]}), and associated uncertainties (dotted lines). Overplotted in red (dashed) is the fit with the exponent fixed at the value observed by \citet{2001MNRAS.324..910C} \textit{i.e.} 1.92. Bottom-right: Inertia-corrected and inertia-frequency-corrected frequency shift $\left[\delta\nu Q_{n,\ell}\epsilon_{n,\ell}\right]$ observed between $\rm Min_{23}$ and $\rm Min_{24}$. Overplotted in black is a fit to the data (solid line), described by Equation (\ref{eq[freq_shift]}), and associated uncertainties (dotted lines). Overplotted in red (dashed) is the fit where $\beta$ was obtained for the $\rm Max_{23}$--$\rm Min_{23}$ comparison. In all plots colours represent different harmonic degrees [$\ell$], as indicated by the colourbar. }
  \label{figure[fit]}
\end{figure}

We model the frequency shift in terms of a power law of the form \citep[\textit{e.g.}][]{1990LNP...367..283G}
\begin{equation}\label{eq[freq_shift]}
\delta\nu=\alpha\nu^\beta,
\end{equation}
where $\delta\nu$ represents the shift in frequency, $\nu$ represents the $m$-averaged frequency, and $\alpha$ and $\beta$ are constants. The value of $\beta$ depends on the nature of the perturbation to the observed frequencies \citep[\textit{e.g.}][]{1990LNP...367..283G, 1990Natur.345..779L, 1991ApJ...370..752G, 2005ApJ...625..548D}. \citet{2001MNRAS.324..910C} observed a change in this relationship at a frequency of $2500\,\rm\mu Hz$ because of a sharp change in the upper turning point of the oscillations at this frequency. We have, therefore, only used modes with frequencies greater than $2500\,\rm\mu Hz$ when fitting the data. We have also set an upper limit of $3500\,\rm\mu Hz$ because the size of the mode frequency uncertainties become increasingly large above this frequency. The fits were made in log space so that the dependence described by equation (\ref{eq[freq_shift]}) corresponds to a straight line. Fits were performed for both the $\rm Min_{23}$--$\rm Min_{24}$ and $\rm Max_{23}$--$\rm Min_{23}$ frequency shifts and the results are plotted in Figure \ref{figure[fit]}. The obtained values of $\alpha$ and $\beta$ are given in Table \ref{table[fits]}.  We note that equation (\ref{eq[freq_shift]}) was fitted to the set containing all of the frequency shifts for all of the individual modes or, in other words, all of the data points plotted in Figure \ref{figure[fit]}.  We initially fitted the inertia corrected frequency shifts given by $\delta\nu_{n,\ell}Q_{n,\ell}$. While both $\alpha$ and $\beta$ are different for the $\rm Min_{23}$--$\rm Min_{24}$ and the $\rm Max_{23}$--$\rm Min_{23}$ fits we note that the value of $\alpha$ obtained for the $\rm Min_{23}$--$\rm Min_{24}$ comparison is poorly constrained. Therefore, as a further test, the $\rm Min_{23}$--$\rm Min_{24}$ data were fitted to determine $\alpha$ only, when $\beta$ was fixed at the value determined for the $\rm Max_{23}$--$\rm Min_{23}$ fit. This fit is also plotted in Figure \ref{figure[fit]}. 

To compare with the results of \citet{2001MNRAS.324..910C} we also fitted frequency shifts with the additional $\epsilon_{n,\ell}$ scaling incorporated, \textit{i.e.} $\delta\nu_{n,\ell}Q_{n,\ell}\epsilon_{n,\ell}$. The results of these fits are given in Table \ref{table[fits]} and are plotted in Figure \ref{figure[fit]}. \citet{2001MNRAS.324..910C} determined an exponent of $1.92\pm0.03$ for GONG modes with $4\le\ell\le150$ and $2500\le\nu\le4000\,\rm\mu Hz$. While this is larger than the exponents that we determined here, Figure \ref{figure[fit]} shows that a value of $\beta=1.92$ is not completely inconsistent with the data. The differences could potentially be explained in terms of different choices over which modes to include in the analysis. Indeed, \citet{2001MNRAS.324..910C} found the determined value of the exponent to depend on which $\ell$ were considered. We again find that the exponent determined for the $\rm Min_{23}$--$\rm Min_{24}$ comparison is consistent with the  $\rm Max_{23}$--$\rm Min_{23}$ comparison (within $2\sigma$) and that the $\rm Max_{23}$--$\rm Min_{23}$ comparison value of $\beta$ also provides a good fit to the $\rm Min_{23}$--$\rm Min_{24}$ data.

\begin{table}\caption{Parameters of power-law fits to frequency shifts, as described by Equation (\ref{eq[freq_shift]}).}\label{table[fits]}
\begin{tabular}{cccc}
  \hline
 Frequency shift & Epochs  & $\alpha$ & $\beta$\\
  \hline
\multirow{3}{*}{$\delta\nu Q_{n,\ell}$} & $\rm Max_{23}$-$\rm Min_{23}$ & $(2.53\pm0.16)\times10^{-14}$ & $3.739\pm0.008$ \\
& $\rm Min_{23}$-$\rm Min_{24}$ & $(6.09\pm0.78)\times10^{-14}$ & $3.353\pm0.072$ \\
& $\rm Min_{23}$-$\rm Min_{24}$ & $(0.27\pm0.03)\times10^{-14}$ & $3.739\pm0.008^*$  \\
  \hline
\multirow{3}{*}{$\delta\nu Q_{n,\ell}\epsilon_{n,\ell}$}  & $\rm Max_{23}$-$\rm Min_{23}$ & $(40.84\pm2.26)\times10^{-8}$ & $1.670\pm0.008$ \\
& $\rm Min_{23}$-$\rm Min_{24}$ & $(2.62\pm0.89)\times10^{-8}$ & $1.718\pm0.042$ \\
& $\rm Min_{23}$-$\rm Min_{24}$ & $(3.85\pm0.02)\time10^{-8}$ & $1.670\pm0.008^*$  \\
  \hline
\end{tabular}\\
$^*$Value in fit was fixed to be the same as the $\rm Max_{23}$-$\rm Min_{23}$ fit.
\end{table}

 \section{Discussion}\label{section[discussion]}
A significant offset is observed in the \textit{p}-mode oscillation frequencies between the minimum preceding $\left[\rm Min_{23}\right]$ and the minimum following $\left[\rm Min_{24}\right]$ Cycle 23, in agreement with, for example, \citet{2010ApJ...711L..84T, 2011ApJ...739....6J}. This suggests that the near-surface magnetic field in the solar interior was weaker during $\rm Min_{24}$. A more detailed analysis of the frequency shifts observed between the two minima suggests that the average shift stops being significant for modes with frequencies below approximately $1800\,\rm\mu Hz$, which implies that the perturbation lies in the upper $0.45$\,percent by radius of the solar interior (or approximately 3\,Mm), based on the upper turning points of the modes \citep{2012ApJ...758...43B}. Using Sun-as-a-star helioseismic data, \citet{2012ApJ...758...43B} determined that the magnetic layer responsible for the solar-cycle dependence of the helioseismic frequencies was thinner in Cycle 23 than Cycle 22. Furthermore, using the upper turning points of the oscillations \citet{2012ApJ...758...43B} inferred that the layer extended over only the outer 0.45 percent by radius. This is comparable to the estimated depth of the magnetic layer perturbation observed here. \citet{2002ApJ...580.1172H} used unsigned-magnetic-flux synoptic maps observed by the Kitt Peak Vacuum Telescope to estimate shift per unit Gauss that a \textit{p}-mode experiences. Table \ref{table[db]} quotes the ``sensitivity values'' determined by \citet{2002ApJ...580.1172H} in their Table 1. Based upon the fit to Equation (\ref{eq[freq_shift]})  we have determined the shift experienced by modes of various frequencies between the two minima and used the sensitivity values of \citet{2002ApJ...580.1172H} to convert this into a change in magnetic field strength. As can be seen from Table \ref{table[db]} this corresponds to an average change of approximately $1\,\rm G$ between the two minima.

\begin{table}\caption{Change in magnetic field [d$B$] as a function of mode frequency. The size of the frequency shift has been determined using Equation (\ref{eq[freq_shift]}) and the values for $\alpha$ and $\beta$ given in Table \ref{table[fits]}. Mode frequencies and sensitivity values [$s_{n\ell}$] giving frequency shift per unit Gauss are taken from Table 1 in \citet{2002ApJ...580.1172H}. }\label{table[db]}
\begin{tabular}{cccc}
  \hline
 Frequency & $\delta\nu Q_{n,l}$ & $s_{n\ell}$  & d$B$ \\
 $[\mu\rm Hz]$ &  [$\mu\rm Hz$] &  [$\rm nHz\,G^{-1}$] & [G] \\
  \hline
2705.81 & $0.019\pm0.003$ & $24.29\pm0.08$ & $0.81\pm0.10$ \\
2892.97 & $0.024\pm0.003$ & $26.65\pm0.07$ & $0.92\pm0.12$ \\
3091.07 & $0.031\pm0.004$ & $29.77\pm0.08$ & $1.03\pm0.13$ \\
3296.25 & $0.038\pm0.005$ & $33.32\pm0.09$ & $1.14\pm0.15$ \\
3498.30 & $0.046\pm0.006$ & $38.83\pm0.14$ & $1.20\pm0.15$\\
  \hline
\end{tabular}\\
\end{table}

The value of $\beta$ in Equation (\ref{eq[freq_shift]}) provides information about the depth dependence of the perturbation responsible for the frequency shifts. A value of 3 is consistent with a change in the speed of sound in a thin layer at the photosphere \citep{1980tsp..book.....C, 1990LNP...367..283G, 1990Natur.345..779L, 1991ApJ...370..752G}.  That the same power law is appropriate to describe both the $\rm Min_{23}$--$\rm Min_{24}$ and $\rm Max_{23}$--$\rm Min_{23}$ frequency shifts indicates that the magnetic field responsible for the perturbation to the frequency shifts was still present in $\rm Min_{23}$, or to put it another way, the near-surface magnetic field was stronger in $\rm Min_{23}$ than in $\rm Min_{24}$.  This raises questions concerning the true baseline magnetic field of the Sun. 

Even at solar minimum, evidence of the Sun's magnetic field can be observed in the form of the ``magnetic carpet''. The magnetic carpet consists of ephemeral regions, intranetwork fields, and network features \citep[see][for a recent review]{2015SSRv..tmp..116C}.  \citet{2011GeoRL..38.6701S} suggested that the Sun approached its global minimum magnetic state in 2008\,--\,2009, still with a population of relatively small dipoles existing. They compare this to the probable state during the Maunder Minimum. Furthermore, \citet{2011GeoRL..38.6701S} find that the minimal magnetic state has a photospheric flux of approximately $1.5\times10^{23}\,\rm Mx$, which corresponds to a mean surface magnetic field of approximately $3\,\rm G$. This means that for an $\ell=0$ \textit{p}-mode with a frequency of approximately $3000\,\rm\mu Hz$ the baseline frequency could be of the order of $0.1\,\rm\mu Hz$ below the frequencies obtained even during $\rm Min_{24}$. We recall that the magnitude of the magnetic activity perturbation is dependent upon $\ell$, or more specifically the mode inertia, and this will impact the difference between the observed and baseline frequencies accordingly. While $0.1\,\rm\mu Hz$ is only a small difference, it is comparable to, and often larger than, the uncertainties associated with the fitted frequencies. Helioseismology could, therefore, provide a vital tool in quantifying low levels of activity in the Sun and determining whether the Sun does have a minimal magnetic state. Furthermore, if, as some have speculated, the Sun is entering an extended period of low activity \citep[\textit{e.g.}][]{2014JGRA..119.3281Z, 2015SoPh..290.1457Z}, helioseismic frequencies may provide a useful measure of small variations in global magnetic field that could still persist throughout an extended magnetically quiet epoch.

In terms of impact on solar and stellar modeling, the amplitude of the frequency shift is still small in comparison to the known surface effects, which cause observational and model \textit{p}-mode frequencies to diverge \citep[\textit{e.g.}][]{1990LNP...367..283G, 2014A&A...568A.123B}. This divergence between models and the observed frequencies is known as the ``surface term'' and is often corrected using a parametric fit \citep[see][and references therein]{2014A&A...568A.123B}. Advances are being made in rectifying these discrepancies \citep{2017MNRAS.464L.124H} and so in the future one may consider what frequencies correspond to a Sun with no magnetic field and the impact of this on standard solar and stellar models. Recently, \citet{2017MNRAS.464.4777H} parameterized solar-cycle variations in the surface term with both a cubic and an inverse frequency term, following the work of \citet{2014A&A...568A.123B}. They found that while the cubic term is more important, for medium-$\ell$ modes the inclusion of the inverse term improved fits to the temporal variation. Therefore, inclusion of an inverse term may be worth considering in future works along the lines presented in this article.

As a slight note of caution, we recall that strong magnetic-flux regions, such as sunspots, do impact \textit{p}-mode frequencies. While the number of visible sunspots is low at solar minimum, as noted in Table \ref{table[proxies]}, there were nevertheless sunspots present during both minimum epochs examined. Furthermore, these records only concern the Earth-facing portion of the Sun: It is possible that farside spots could also be present. A future study may consider isolating periods where no spots were observed on the solar surface. This would require moving away from the 108-day bins upon which the GONG frequencies are based. Comparison with observed frequencies as systematically increased numbers of sunspots appear would, however, provide insights into the proportion of the global \textit{p}-mode frequency shifts that could be accounted for by sunspots.

Finally, we note that we have deliberately avoided calling the minimum following Cycle 23 peculiar because this article only examines two minima and there is evidence for similar minima in historical records \citep[\textit{e.g.}][]{2010ASPC..428..171J, 2010ASPC..428....3S}. It will be interesting to follow the upcoming minimum as this may provide evidence both for and against the Sun entering a Maunder-Minimum-like state. Regardless, this article highlights the increased importance of maintaining coherent observations through the approaching solar minimum and beyond. 

%
\begin{acks}
We thank the anonymous referee, whose comments have substantially improved the paper. A.-M. Broomhall thanks the Institute of Advanced Study, University of Warwick for their support. She also thanks Y. Elsworth and V.M. Nakariakov for their insightful comments. This work utilizes GONG data obtained by the NSO Integrated Synoptic Program (NISP), managed by the National Solar Observatory, which is operated by the Association of Universities for Research in Astronomy (AURA), Inc. under a cooperative agreement with the National Science Foundation. The data were acquired by instruments operated by the Big Bear Solar Observatory, High Altitude Observatory, Learmonth Solar Observatory, Udaipur Solar Observatory, Instituto de Astrof\'isica de Canarias, and Cerro Tololo Interamerican Observatory. This study includes data from the synoptic program at the150-Foot Solar Tower of the Mt. Wilson Observatory. The Mt. Wilson 150-Foot Solar Tower is operated by UCLA, with funding from NASA, ONR, and NSF, under agreement with the Mt. Wilson Institute. We thank NOAA NGDC and the Royal Observatory (Greenwich) for making their data freely available. Wilcox Solar Observatory data used in this study was obtained via the web site \href{http://wso.stanford.edu}{\textsf{wso.stanford.edu}} at 01:00 PDT on 12 August 2016 courtesy of J.T. Hoeksema. The Wilcox Solar Observatory is currently supported by NASA. We thank PMOD/WRC, Davos, Switzerland of the updated dataset (\textsf{$\sf composite\_42\_65\_1605.dat$}) with new data from the VIRGO Experiment on the cooperative ESA/NASA Mission SOHO.
\end{acks} 

\section*{Disclosure of Potential Conflicts of Interest} The author declares that she has no conflicts of interest.

\bibliographystyle{spr-mp-sola}
\bibliography{minima} 

\end{article}
\end{document}